\documentclass[twocolumn,epjc3]{svjour3}          

\usepackage{graphicx}
\usepackage{dcolumn}
\usepackage{bm}
\usepackage{amsmath,multirow}
\usepackage{amssymb}
\usepackage{subfigure}
\usepackage{color}

\RequirePackage[T1]{fontenc}

\smartqed  

\journalname{EPJC}

\begin{document}

\title{Applying explicit
symplectic integrator to study chaos of charged particles around
magnetized Kerr black hole }


\author{
Wei Sun$^{1,2}$ \and
Ying Wang$^{1,2}$ \and
Fuyao Liu$^{1}$ \and
Xin Wu$^{1,2,3}$} 

\thankstext{e2}{e-mail: wuxin$\_$1134@sina.com}

\institute{School of Mathematics, Physics and Statistics, Shanghai
University of Engineering Science, Shanghai 201620, China
\label{addr1}
          \and Center of Application and Research of Computational Physics,
Shanghai University of Engineering Science, Shanghai 201620, China
\label{addr2}
          \and Guangxi Key Laboratory for Relativistic Astrophysics, Guangxi
University, Nanning 530004, China
}

\date{Received: date / Accepted: date}

\maketitle

\begin{abstract}

In a recent work of Wu, Wang, Sun and Liu, a second-order explicit
symplectic integrator was proposed for the integrable Kerr
spacetime geometry. It is still suited for simulating the
nonintegrable dynamics of charged particles moving around the Kerr
black hole embedded in an external magnetic field. Its successful
construction is due to the contribution of a time transformation.
The algorithm exhibits a good long-term numerical performance in
stable Hamiltonian errors and computational efficiency.  As its
application, the dynamics of order and chaos of charged particles
is surveyed. In some circumstances, an increase of the dragging
effects of the spacetime seems to weaken the extent of chaos from
the global phase-space structure on Poincar\'{e} sections.
However, an increase of the magnetic parameter strengthens the
chaotic properties. On the other hand, fast Lyapunov indicators
show that there is no universal rule for the dependence of the
transition between different dynamical regimes on the black hole
spin. The dragging effects of the spacetime do not always weaken
the extent of chaos from a local point of view.

\end{abstract}



\section{Introduction}
\label{sec:intro}

The Kerr spacetime [1] as well as the Schwarzschild spacetime is
integrable because the Carter constant [2] appears as a fourth
constant of motion. However, a test charge motion in a magnetic
field around the Kerr black hole or the Schwarzschild black hole
is nonintegrable due to the absence of the fourth constant of
motion. In some cases, the electromagnetic perturbation induces
the onset of chaos and islands of regularity [3-10]. In
particular, Takahashi $\&$ Koyama [11] found the connection
between chaoticness of the motion and the black hole spin. That
is, the chaotic properties are weakened with an increase of the
black hole spin.

It is worth pointing out that the detection of the chaotical
behavior needs reliable results on the trajectories. Usually, it
needs long-time calculations, especially in the case of the
computation of Lyapunov exponents. Thus, the adopted computational
scheme is required to have good stability, high precision and
small cost of computational time. Because these general
relativistic curved spacetimes correspond to Hamiltonian systems
with the symplectic nature, a low order symplectic integration
scheme  which respects the symplectic nature  of Hamiltonian
dynamics [12-14] is naturally regarded to as the most appropriate
solver. The Hamiltonian systems for these curved spacetimes are
not separable to the phase-space variables, therefore, no explicit
symplectic integrators but implicit (or implicit and explicit
combined) symplectic integrators [15-22] can be available in
general. However, the implicit methods are more computationally
demanding than the explicit ones at same order. To solve this
problem, Wang et al. [23-25] split the Hamiltonians of
Schwarzschild, Reissner-Nordstr\"{o}m and Reissner-Nordstr\"{o}m
-(anti)-de Sitter spacetime geometries into four, five and six
integrable separable parts, whose analytical solutions are
explicit functions of proper time. In this way, explicit
symplectic integrators were designed for the spacetime geometries.
Unfortunately, the dragging effects of the spacetime by a rotating
black hole make the Hamiltonian of Kerr spacetime have no desired
splitting form similar to the Hamiltonian splitting form of
Schwarzschild spacetime, and then lead to the difficulty in the
construction of explicit symplectic integrators. This obstacle was
successfully overcame by the introduction of time transformations
of Mikkola [26] to the Hamiltonian of Kerr spacetime geometry. The
obtained time-transformed Hamiltonian exists a desired splitting,
and explicit symplectic integrators were established by Wu et al.
[27]. At this point, a note is worthwhile on such time-transformed
explicit symplectic integrators. They are constant time-steps in
the new coordinate time and the algorithmic symplecticity can be
maintained. Although they are variant time-steps in the proper
coordinate time, the varying proper time-steps have only small
differences and are approximately constant when the new
integration time is not very long. In other words, the time
transformation plays a main role in obtaining separable
time-transformed Hamiltonian for the use of explicit symplectic
integrators rather than adaptive time step control.

One of the main purposes in this paper is to explore whether the
proposed explicit symplectic integrators for the  Kerr spacetime
geometry work well in the simulation of the chaotic motion of
charged particles near the Kerr black hole immersed in a weak,
asymptotically uniform external magnetic field. As another
important purpose, we apply the established second-order explicit
symplectic method to check the result of Takahashi $\&$ Koyama
[11] on the dragging effects weakening the chaotic properties, and
to survey how a small change of the magnetic parameter affects a
dynamical transition from order to chaos. Besides the Poincar\'{e}
map method, the methods of Lyapunov exponents [28, 29] and fast
Lyapunov indicators [30-32] are employed to distinguish between
regular and chaotic orbits and to quantify the dependence of
transition from regular to chaotic dynamics on a certain
parameter.

For the sake of our purposes, we introduce a Hamiltonian system
for the description of charged particles moving around the Kerr
black hole surrounded with an external magnetic field in Sect. 2.
A second-order time-transformed explicit symplectic algorithm is
constructed and used to quantify the transition from regular to
chaotic dynamics by means of different methods finding chaos in
Sect. 3. Finally, the main results are concluded in Sect. 4.

\section{Kerr black hole with external magnetic field}

Let us consider a particle with charge $q$ moving around the Kerr
black hole, surrounded by an external asymptotically uniform
magnetic field with strength $B$. The particle motion is described
by the following Hamiltonian formalism [33]
\begin{equation}
H = \frac{1}{2}g^{\mu\nu}(p_{\mu}-qA_{\mu})(p_{\nu} -qA_{\nu}).
\end{equation}
Contravariant Kerr metric $g^{\mu\nu}$  has nonzero components
\begin{eqnarray}
&& g^{tt}=-\frac{D}{\Lambda\Sigma}, ~~
g^{t\phi}=-\frac{2ar}{\Lambda\Sigma}=g^{\phi t}, \nonumber \\
&&  g^{rr}=\frac{\Lambda}{\Sigma}, ~~~~~~
g^{\theta\theta}=\frac{1}{\Sigma},  \nonumber \\
&& g^{\phi\phi}=\frac{\Sigma-2r}{\Lambda\Sigma\sin^{2}\theta};
\nonumber \\
&& \Sigma=r^2+a^2\cos^{2}\theta, ~~~~\Lambda=r^2+a^2-2r, \nonumber \\
&& D=(r^2+a^2)^2-\Lambda a^2\sin^{2}\theta. \nonumber
\end{eqnarray}
$A^{\mu}$ is an electromagnetic four-vector potential [33, 34]
\begin{equation}
A^{\mu}=aB\xi^{\mu}_{(t)}+\frac{B}{2}\xi^{\mu}_{(\phi)},
\end{equation}
where $\xi^{\mu}_{(t)}=(1,0,0,0)$ and
$\xi^{\mu}_{(\phi)}=(0,0,0,1)$ are timelike and spacelike axial
Killing vectors. Obviously, the four-vector potential has two
nonzero covariant components
\begin{eqnarray}
A_{t} &=& aBg_{tt}+\frac{B}{2}g_{t\phi}, \nonumber \\
&=& -aB[1+\frac{r}{\Sigma}(\sin^{2}\theta-2)] \\
A_{\phi} &=& aBg_{t\phi}+\frac{B}{2}g_{\phi\phi} \nonumber \\
&=& B\sin^{2}\theta[\frac{r^2+a^2}{2}
+\frac{a^2r}{\Sigma}(\sin^{2}\theta-2)].
\end{eqnarray}
The above  covariant metric's components are
\begin{eqnarray}
&& g_{tt}=-(1-\frac{2r}{\Sigma}), ~~
g_{t\phi}=-\frac{2ar\sin^{2}\theta}{\Sigma}=g_{\phi t}, \nonumber \\
&&  g_{\phi\phi}=(r^2+a^2+\frac{2ra^2}{\Sigma}
\sin^{2}\theta)\sin^{2}\theta. \nonumber
\end{eqnarray}

$p_{\mu}$ represents a covariant generalized four-momentum,
determined by the relation
\begin{equation}
\dot{x}^{\mu}=\frac{\partial H}{\partial p_{\mu}}
=g^{\mu\nu}(p_{\nu} -qA_{\nu}).
\end{equation}
Note that $\dot{x}^{\mu}$ is a derivative of coordinate
$x^{\mu}=(t,r,\theta,\phi)$ with respect to proper  time $\tau$,
called as a 4-velocity. Eq. (5) is rewritten as
\begin{equation}
p_{\mu}= g_{\mu\nu}\dot{x}^{\nu}+qA_{\mu}.
\end{equation}
Because $\dot{p}_{\mu}=-\partial H/\partial x^{\mu}$, $p_{t}$ and
$p_{\phi}$ are constants of motion. They are the particle's energy
and angular momentum:
\begin{eqnarray}
-E &=& p_{t}= g_{tt}\dot{t}+g_{t\phi}\dot{\phi} +qA_{t}, \\
L &=& p_{\phi}= g_{t\phi}\dot{t}+g_{\phi\phi}\dot{\phi}
+qA_{\phi}.
\end{eqnarray}
They have equivalent expressions
\begin{eqnarray}
\dot{t} &=& -f_1(E+qA_t)-f_2(L-qA_{\phi}), \\
\dot{\phi} &=& f_2(E+qA_t)+f_3(L-qA_{\phi}),
\end{eqnarray}
where
\begin{eqnarray}
f_1 &=& \frac{g_{\phi\phi}}{g_{tt}g_{\phi\phi}-g^{2}_{t\phi}}, \\
f_2 &=& \frac{g_{t\phi}}{g_{tt}g_{\phi\phi}-g^{2}_{t\phi}}, \\
f_3 &=& \frac{g_{tt}}{g_{tt}g_{\phi\phi}-g^{2}_{t\phi}}.
\end{eqnarray}

Now, Eq. (1) is rewritten as
\begin{eqnarray}
H &=& \frac{1}{2}g^{\mu\nu}(p_{\mu}-qA_{\mu})(p_{\nu}
-qA_{\nu}) \nonumber \\
&=& F +\frac{1}{2}\frac{\Lambda}{\Sigma}p^{2}_{r}
+\frac{1}{2}\frac{p^{2}_{\theta}}{\Sigma},
\end{eqnarray}
where $F$ is a function of $r$ and $\theta$ as follows:
\begin{eqnarray}
F &=&\frac{1}{2}[g^{tt}(E+qA_{t})^2 +g^{\phi\phi}(L-qA_{\phi})^2]\nonumber \\
&&  -g^{t\phi}(E+qA_{t})(L-qA_{\phi}) \nonumber \\
&=& \frac{1}{2}[f_1(E+qA_{t})^2 +f_3(L-qA_{\phi})^2] \nonumber \\
&&  +f_2(E+qA_{t})(L-qA_{\phi}) \nonumber \\
&=&
-\frac{D}{2\Lambda\Sigma}(E+qA_{t})^2+\frac{\Sigma-2r}{2\Lambda
\Sigma\sin^{2}\theta}(L-qA_{\phi})^2 \nonumber \\
&& +\frac{2ar}{\Lambda\Sigma}(E+qA_{t})(L-qA_{\phi}).
\end{eqnarray}

In Eq. (14), we take the speed of light $c$ and the gravitational
constant $G$ as geometrized units, $c=G=1$. The black hole's mass
$M$ is also one unit, $M=1$. In practice, dimensionless operations
are given to the related quantities via a series of scale
transformations. That is, $r\rightarrow rM$, $t\rightarrow tM$,
$\tau\rightarrow \tau M$, $a\rightarrow aM$, $E\rightarrow Em$,
$p_r\rightarrow mp_r$, $L\rightarrow mML$, $p_{\theta}\rightarrow
mMp_{\theta}$, $q\rightarrow mq$, $B\rightarrow B/M$ and
$H\rightarrow m^2H$, where $m$ denotes the particle's mass.
Hereafter, we take $\beta=qB$. For any rotating black hole in
general relativity, its spin angular momentum always satisfies
$|a|\leq 1$.

Because the timelike 4-velocity (5) always satisfies the relation
$\dot{x}^{\mu}\dot{x}_{\mu}=-1$, the Hamiltonian (14) itself is
identical to a given constant
\begin{equation}
H=-\frac{1}{2}.
\end{equation}
Besides this value and $E$ and $L$, other constants do not exist
in the system. Thus, the Hamiltonian is non-integrable.

\section{Numerical investigations}

Following the work of Wu et al. [27], we establish a second-order
explicit symplectic integrator for a time-transformed Hamiltonian
of the system (14). For comparison, a second-order implicit and
explicit mixed symplectic method is applied to solve the
Hamiltonian (14) in Sect. 3.1. Then, the explicit symplectic
integrator is used to investigate the dependence of the regular
and chaotic dynamics of the time-transformed Hamiltonian on a
certain parameter by means of several methods finding chaos in
Sect. 3.2.

\subsection{Numerical integration scheme}

Clearly, the Hamiltonian (14) is inseparable to the variables.
Implicit symplectic methods can be applied to it without doubt.
The first term $F$ in Eq. (14) is solved analytically, and the sum
of the second and third terms is solved numerically by  the
second-order implicit midpoint symplectic method IM2 [15]. The
explicit and implicit solutions compose a second-order implicit
and explicit mixed symplectic integrator for  $H$, labeled as IE2.
Such a mixed integrators was discussed by several authors [17-22].
It can save labour compared with the midpoint rule IM2 directly
acting on the whole Hamiltonian $H$.

On the other hand, Wu et al. [27] used a time transformation
function
\begin{equation}
d\tau=g(r,\theta)dw, ~~ g(r,\theta)=\frac{\Sigma}{r^2}
\end{equation}
to obtain a time-transformed Hamiltonian to the Kerr geometry. The
time-transformed Hamiltonian can be solved by explicit symplectic
integrators. In the present problem, the time-transformed
Hamiltonian is
\begin{equation}
K=g(H+p_0)=\frac{\Sigma}{r^2} (F+p_0)
+\frac{\Lambda}{2r^2}p^{2}_{r} +\frac{1}{2r^2}p^{2}_{\theta}.
\end{equation}
$p_0=-H=1/2$ is a momentum with respect to coordinate $q_0=\tau$.
Thus, $K$ is always identical to zero, $K=0$. The time-transformed
Hamiltonian is split into five parts
\begin{equation}
K=K_1+K_2+K_3+K_4+K_5,
\end{equation}
where the sub-Hamiltonians read
\begin{eqnarray}
K_1 &=& \frac{\Sigma}{r^2} (F+p_0), \\
K_{2} &=& \frac{1}{2}p^{2}_{r},\\
K_{3} &=& -\frac{1}{r}p^{2}_{r},\\
K_{4} &=&
\frac{a^2}{2r^2}p^{2}_{r}, \\
K_{5} &=& \frac{1}{2r^2}p^{2}_{\theta}.
\end{eqnarray}
$K_{1}$ seems to be the same as that of Wu et al. [27] from the
expressional form, but is unlike that of Wu et al. $K_{2}$,
$K_{3}$, $K_{4}$ and $K_{5}$ are the same as those of Wu et al.
The five sub-Hamiltonians have analytical solutions that are
explicit functions of the new coordinate time $w$. Their solutions
correspond to operators $\widetilde{K}_{1}$, $\widetilde{K}_{2}$,
$\widetilde{K}_{3}$, $\widetilde{K}_{4}$ and $\widetilde{K}_{5}$.
According to the idea of Wu et al., these operators can compose
second-order explicit symplectic integrator for  $K$:
\begin{eqnarray}
S2(h) &=& \widetilde{K}_5(\frac{h}{2})\circ
\widetilde{K}_4(\frac{h}{2})\circ
\widetilde{K}_3(\frac{h}{2})\circ
\widetilde{K}_2(\frac{h}{2})\circ \widetilde{K}_1(h) \nonumber
\\ && \circ \widetilde{K}_2(\frac{h}{2})\circ
\widetilde{K}_3(\frac{h}{2})\circ \widetilde{K}_4(\frac{h}{2})
\circ \widetilde{K}_5(\frac{h}{2}),
\end{eqnarray}
where $h$ represents a new coordinate time-step. The use of fixed
new coordinate time-step maintains the symplecticity of the
algorithms for the time-transformed Hamiltonian $K$. However, the
proper time $\tau$ takes varying step-sizes. Such symplectic
integrators are adaptive time-steps [26, 35, 36]. However, the
adaptive proper time step control is almost absent in the present
problem because the time transformation function $1<g\leq
1+1/r^2\approx 1$ for $r\gg1$. This indicates that the time
transformation function mainly provides the desirable separable
time-transformed Hamiltonian.

Based on the numerical results of Wu et al., the second-order
explicit symplectic method S2 performs good computational
efficiency and stabilizing error behavior, compared with the
fourth-order explicit symplectic method S4. Thus, we focus on the
application of S2 to  the system (19). For comparison, the
second-order implicit and explicit mixed symplectic integrator IE2
is applied to  $H$. Let us take the time-step $h=1$. Note that
$h=1$ denotes a proper time-step for IE2 and a new coordinate
time-step for S2. The parameters are given by $E=0.995$, $L=4.6$,
$a=0.5$ and $\beta=0.001$. The initial conditions are
$\theta=\pi/2$ and $p_r=0$. The initial separations $r_0=11$ and
75 are respectively given to Orbits 1 and 2. The initial value of
$p_{\theta}>0$ is determined by Eqs. (14) and (16). It is shown in
Fig. 1(a) that Hamiltonian errors $\Delta =K$ for the explicit
symplectic algorithm S2 and $\Delta =-H-1/2$ for the second-order
implicit and explicit  mixed symplectic method IE2 remain stable
and bounded. Both algorithms give no explicit error differences to
the same orbit. The errors have the same order for the two orbits.
If IE2 is applied to $K$, the results are almost the same as those
for the application of IE2 to $H$. Although S2 is not explicitly
better than IE2 in numerical accuracy, it is in computational
efficiency, as claimed in the work of Wu et al.

It is worth pointing out that IE2 and S2 work in different time
systems.  The integration time for IE2 is the proper time
$\tau=10^{8}$. However, the integration time for S2 is the new
coordinate time $w=10^{8}$. The new coordinate time $w=10^{8}$ and
its corresponding real proper time have no large differences. This
result is shown clearly in Table 1. In fact, differences between
the solutions of $H$ and those of $K$ are negligible before the
integration time arrives at $10^{5}$. So are the differences
between the proper times of $K$ and the coordinate times of $K$.
These facts confirm that the time transformation does not provide
adaptive time-steps but a separable time-transformed Hamiltonian.
Of course, these differences between the  solutions of the two
systems get slowly large as the integration time spans $10^{5}$
and increases. In this case, the solutions of $K$ at a given
proper time are obtained with the aid of some interpolation
method.

In fact, the tested orbits have different dynamical behaviors, as
shown on Poincar\'{e} sections at the plane $\theta=\pi/2$ and
$p_{\theta}>0$ in Fig. 1(b). The phase-space structure of Orbit 1
is a torus, regarded as the characteristic of an ordered orbit.
However, the phase-space structure of Orbit 2 has many discrete
points that are randomly filled in an area, regarded as the
characteristic of a chaotic orbit. It can be seen from Fig. 1(a)
that the performance of S2 (with IE2) is independent of the
regularity and chaoticity of orbits. The phase-space structures in
Fig. 1(b) are described by S2. They are also given by IE2.

\subsection{Dynamical transition with parameters varying}

In what follows, we use the algorithm S2 to trace the orbital
dynamics of the system $K$. Meantime, methods of Poincar\'{e}
sections, Lyapunov exponents and fast Lyapunov indicators are
employed to find chaos.

\subsubsection{Poincar\'{e} sections}

It is worth pointing out that the chaoticity of Orbit 2 is due to
the charged particle suffered from the electromagnetic  field
interaction. As claimed in the Introduction, the Kerr spacetime
without the electromagnetic field interaction holds the Carter
constant as the fourth integral of motion and therefore is
integrable and nonchaotic. If the electromagnetic field is
included in the Kerr spacetime, then the Carter constant is no
longer present and the dynamics of charged particles becomes
nonintegrable. This nonintegrability is an insufficient but
necessary condition for the occurrence of chaos. When the
electromagnetic  field interaction acts as a relatively small
perturbation, the gravitational force from the black hole is a
dominant force and the dynamical system is nearly integrable. In
this case, the Kerr spacetime with the electromagnetic field
interaction can exhibit the similar dynamical features of the Kerr
spacetime without the electromagnetic field interaction. That is
to say, no chaos occurs and all orbits are regular
Kolmogorov-Arnold-Moser (KAM) tori. This result is suitable for
the magnetic parameter $\beta=4\times 10^{-4}$ in Fig. 2(a). The
tori in the perturbed case unlike those in the unperturbed case
are twisted in the shapes; e.g., Orbit 1 corresponds to a peculiar
torus, and Orbit 4 with initial separation $r_0=110$ yields a
triangle torus.  When $\beta=6\times 10^{-4}$ in Fig. 2(b),  the
tori are still present, but are to a large degree different from
those for $\beta=4\times 10^{-4}$ in the shapes. There are typical
differences of Orbits 1, 2 and 3 between Figs.  2 (a) and (b). In
particular, Orbit 4 in Fig. 2(a) is evolved to a twisted
figure-eight orbit with a hyperbolic point in Fig. 2(b) due to the
electromagnetic field interaction. Such a hyperbolic point has a
stable  direction and another unstable direction, and easily
induces chaos. As the electromagnetic field interaction increases
and can  generally match with the black hole gravitational force,
chaos occurs. The hyperbolic point causes Orbit 4 to be chaotic
for $\beta=7\times 10^{-4}$ in Fig. 2(c). When the magnetic field
strength is further enhanced, e.g., $\beta=8\times 10^{-4}$ in
Fig. 2(d), $\beta=1\times 10^{-3}$ in Fig. 1(b) and
$\beta=1.2\times 10^{-3}$ in Fig. 2(e), chaos becomes stronger and
stronger. Particular for $\beta=3\times 10^{-3}$ in Fig. 2(f),
chaos is existent almost everywhere in the whole phase space. In a
word, it can be concluded clearly from Figs. 1(b) and 2 that an
increase of the magnetic parameter is easier to induce chaos and
enhances the strength of chaos from the global phase-space
structure.

On the other hand, we are also interested in knowing what the
dynamical transition with an increase of the black hole angular
momentum $a$ is. To answer this question, we consider some values
of $a$. For $a=0$ corresponding to the Schwarzschild case in Fig.
3(a), chaos densely exists almost everywhere in the whole phase
space. This does not mean that any regular orbits are ruled out.
In fact, regular Orbit 2 and black orbit with initial separation
$r_0=25$ in Fig. 1(b) are still ordered in Fig. 3(a). Given
$a=0.2$, chaotic Orbits 1 and 3 in Fig. 3(a) become regular  in
Fig. 3(b), whereas ordered Orbit 2 in Fig. 3(a) becomes chaotic in
Fig. 3(b). As the spin frequently increases [e.g., $a=0.5$ in Fig.
1(b), $a=0.8$ in Fig. 3(c) and $a=1$ in Fig. 3(d)], the regular
region seems to gradually increase and the chaotic region seems to
decrease. In other words, an increase of $a$ seems to weaken the
extent of chaos from the global phase-space structure. This seems
to support the result of [11] about the black hole spin weakening
the chaotic properties. However, this does not mean that some
individual orbits must be more chaotic as the dragging effects of
the spacetime by a rotating black hole decrease. For example,
Orbit 2 is regular for $a$=0, 1 but chaotic for $a$=0.2, 0.5, 0.8.

\subsubsection{Fast Lyapunov indicators}

Apart from the Poincar\'{e} map method, the technique of Lyapunov
exponents is very efficient to detect the onset of chaos and
provides a quantitative measure to chaos. An average exponential
deviation of two nearby orbits in the system $K$ is measured by
the largest Lyapunov exponent
\begin{equation}
\lambda=\lim_{w\rightarrow\infty}\frac{1}{w}\ln\frac{d(w)}{d(0)},
\end{equation}
where $d(0)$ and $d(w)$ represent the separations between the two
nearby orbits at times 0 and $w$, respectively. Precisely
speaking, $d(w)$ is written as
\begin{eqnarray}
d(w) &=& [(r_2-r_1)^2+(\theta_2-\theta_1)^2+(p_{r2}-p_{r1})^2
\nonumber \\
&& +(p_{\theta2}-p_{\theta1})^2]^{1/2},
\end{eqnarray}
where $(r_1,\theta_1,p_{r1},p_{\theta1})$ is a set of phase-space
variables of one orbit at time $w$ and
$(r_2,\theta_2,p_{r2},p_{\theta2})$ is a set of phase-space
variables of its nearby orbit. Three points should be noted. (i)
The initial distance $d(0)$ had better be $10^{-8}$ in the
double-precision case. (ii) Appropriate renormalizations to the
distance $d(w)$ are required from time to time. (iii) Although the
Lyapunov exponent depends on the choice of spacetime coordinates,
it is very approximate to an invariant Lyapunov exponent because
the new coordinate time $w$ and its corresponding proper time
$\tau$ are approximately equal, and the distance $d(w)$ and its
corresponding proper distance are, too. More details about the
invariant Lyapunov exponent were given by Wu $\&$ Huang [29] and
Wu et al. [32]. A positive Lyapunov exponent indicates the
chaoticity of a bounded orbit, and zero Lyapunov exponent
describes the regularity of a bounded orbit. This allows us to
detect chaos from order. It is clearly seen from the Lyapunov
exponents in Fig. 4(a) that Orbits 1 and 2 in Fig. 1 are ordered
and chaotic, respectively. Orbit 3 in Fig. 2(d) seems to be
regular till $w=10^{7}$ because its Lyapunov exponent like Orbit
1's Lyapunov exponent decreases and does not tend to a stabilizing
value. However, the Lyapunov exponent of Orbit 3 unlike that of
Orbit 1 tends to a stabilizing positive value when the integration
time spans from $10^{7}$ to $10^{8}$; therefore, Orbit 3 is
chaotic. Because the Lyapunov exponent of Orbit 3 is smaller than
that of Orbit 2, the chaoticity of Orbit 3 is weaker than that of
Orbit 2. The result is consistent with that described by the
method of Poincar\'{e} sections in Figs. 1(b) and 2(d).

In general, a long enough integration time is necessary to obtain
a stabilizing value of Lyapunov exponent. In particular, the
chaoticity of Orbit 3 cannot be distinguished from the regularity
of Orbit 1 in terms of Lyapunov exponents until $w=10^{8}$.
Compared with the method of Lyapunov exponents, the fast Lyapunov
indicator (FLI) of Froeschl\'{e} $\&$ Lega [31] is based on
computations of tangent vectors and is regarded as a quicker
technique to separate chaotic orbits from regular ones. Wu et al.
[32] modified the FLI with tangent vectors as the FLI of two
nearby orbits:
\begin{equation}
FLI=\log_{10}\frac{d(w)}{d(0)}.
\end{equation}
Here, $d(0)=10^{-9}$ is admissible in the double-precision case,
and a few but not many renormalizations to the distance $d(w)$ are
still necessary to avoid the saturation of orbits. Regular and
chaotic orbits can be identified according to  completely
different time rates of growth of FLIs. An exponential increase of
FLI of a bounded with time $\log_{10} w$ indicates the chaoticity
of the orbit; an algebraical increase of FLI of a bounded with
time $\log_{10} w$ indicates the regularity of the orbit. In other
words, if the FLI of a bounded orbit is much larger than that of
another bounded orbit for a given time, the former orbit is
chaotic, but the latter orbit is regular. On the basis of this
point, the properties of orbits in Fig. 4(b) can be known easily.
Without doubt, Orbit 2 is strongly chaotic because its FLI arrives
at 12 when the integration time $w=6\times 10^5$, and is larger
than 200 when $w=10^7$. Unlike the Lyapunov exponents, the FLIs
are easy to distinguish between the regularity of Orbit 1 and  the
chaoticity of Orbit 3 when $w=10^7$ because the FLI of Orbit 1 is
smaller than 6, but the FLI of Orbit 3 is 13.5.

As claimed in Ref. [32], the FLI is a good tool to allow us to
identify the transition between different dynamical regimes with a
variation of a certain dynamical parameter. Letting $a=0.5$ be
fixed and $\beta$ run from 0 to $1.5\times10^{-3}$, we trace the
dependence of the dynamical transition on the magnetic parameter.
For a given value $\beta$, the integration lasts to the time
$w=10^7$, but it ends if the FLI of some orbit is 20. There are
three areas: regular area with FLIs smaller than 6, weak chaos
area with  FLIs no less than 6 but no more than 20 and strong
chaos area with FLIs larger than 20. When
$\beta<0.7\times10^{-3}$, no chaos occurs in Figs. 5 (a) and (b).
This is because the magnetic field force to charged particles is
smaller than the black hole gravity to charged particles, as is
mentioned above. $\beta<0.7\times10^{-3}$ is a regular area of the
magnetic parameter. Chaos will get easier to a large degree as
$\beta$ increases. When $\beta>0.8\times10^{-3}$, many values of
$\beta$ correspond to the onset of chaos and many other values
indicate the presence of order in Fig. 5(a). In particular, chaos
and order are present in the neighbourhood of
$\beta=1.5\times10^{-3}$. Unlike in Fig. 5(a), all values of
$\beta>1.1\times10^{-3}$ can induce strong chaos in Fig. 5(b). Of
course, few values of the magnetic parameter correspond to weak
chaos in the two panels. In short, the technique of FLIs and the
method of Poincar\'{e} sections consistently show that an increase
of the magnetic parameter leads to a strong probability for
inducing the occurrence of chaos and strengthening the extent of
chaos.

Figs. 6 (a)-(d) plot the dependence of FLIs on the black hole spin
$a$ when $\beta=0.001$ is fixed. Strong chaos occurs as
$a\rightarrow 0$ but no chaos appears as $a\rightarrow 1$ for the
initial separation $r_0=40$ in Fig. 6(a). For the most values of
$a\in[0.25,0.65]$, chaos is absent, too. The transition has some
differences for the initial separation $r_0=60$ in Fig. 6(b).
Chaos exists when $a\rightarrow 1$. However, there is no chaos at
the left end $a=0$ and the right end $a=1$ in Figs. 6 (c) and (d).
These facts show that different combinations of initial conditions
and other dynamical parameters affect the dependence of the
dynamical transition on the black hole spin. Thus, no universal
rule can be given to the relation between the dynamical transition
and the black hole spin.

A notable point is that the aforementioned results are completely
based on mathematical calculations. To justify the mathematical
correctness of the obtained results, we employ an eighth- and
ninth-order Runge-Kutta-Fehlberg integrator (RKF89) with adaptive
step sizes to integrate the system (14) or (19). This algorithm
can provide a machine precision to the Hamiltonians if roundoff
errors are ignored. Of course, the higher-order method is more
expensive in computations than the second-order explicit scheme
S2. Consequently, the results obtained from RKF89 are consistent
with those given by S2. It is very perfect and idealized to make a
comparison of the obtained theoretical results with real
observational results  in the relativistic astrophysics. However,
the known observational results have not sufficiently supported
the study of chaotic behavior of black holes. There are two main
reasons. On one hand, it costs a long enough time span to
distinguish between the regular and chaotic  two cases. The time
for the determination of chaos is $10^7M$ or $10^8M$ from the
numerical computations. Nevertheless, it is impossibly available
from the observations. On the other hand, it costs massive data to
detect chaos from order. This fact can be seen clearly from the
numerical simulations. However, only a small amount of
observational data are given. Because of these reasons, the study
of chaos and islands of regularity of black holes in all the known
publications such as [3-11, 23-25, 33 and 34] is restricted to the
theoretically numerical results rather than the real observational
results.

\section{Conclusion}

Because of the dragging effects of the spacetime by a rotating
black hole, the Kerr geometry is complicated. In spite of this, it
is integrable and nonchaotic. When an external magnetic field is
included, the dynamics of charged particles becomes more
complicated. In fact, it is nonintegrable and possibly chaotic. We
find that the second-order explicit symplectic integrator for the
integrable Kerr spacetime proposed in the work of Wu et al. [27]
is still suited for the present nonintegrable problem. The
successful application of the explicit symplectic integrator is
owing to the contribution of the time transformation given in Eq.
(17). The theoretical analysis and numerical results show that
such a time transformation does not mainly bring adaptive proper
time steps but a separable time-transformed Hamiltonian, which
satisfies a need for the use of explicit symplectic integrator.
The explicit symplectic algorithm performs good performance in
numerical accuracy, stable error behaviour and computational cost
regardless of whether tested orbits are regular or chaotic.

The explicit symplectic integrator is very suitable for studying
the long-term qualitative evolution of orbits of charged particles
moving around the magnetized Kerr black hole. The electromagnetic
force interactions are mainly responsible for the nonintegrability
and chaoticity of charged particle dynamics. In some
circumstances, charged particle motions can be chaotic. An
increase of the dragging effects of the spacetime seems to weaken
the extent of chaos from the global phase-space structure on
Poincar\'{e} sections. However, an increase of the magnetic
parameter results in strengthening the chaotic properties. On the
other hand, the fast Lyapunov indicators by scanning the black
hole spin parameter show that  no universal rule can be given to
the dependence of the transition between different dynamical
regimes on the black hole spin. This is because the transition is
dependent on not only the spin parameter but also different
combinations of initial conditions and other parameters. Unlike an
increase of the black hole spin, that of the magnetic parameter
may much easily induce chaos.

\textbf{Acknowledgements}: The authors are very grateful to two
referees for valuable comments and suggestions. This research has
been supported by the National Natural Science Foundation of China
[Grant Nos. 11973020 (C0035736), 11533004, 11803020, 41807437,
U2031145], and the Natural Science Foundation of Guangxi (Grant
Nos. 2019JJD110006 and 2018GXNSFGA281007).

\begin{table*}[htbp]
\centering \caption{Solutions for the methods S2 and IE2 solving
Orbit 1 in Fig. 1(a). The two systems $H$ and $K$ almost have the
same solutions in an integration time of $\tau=10000$ or
$w=10000$. } \label{Tab1}
\begin{tabular}{lcccccccccccc}
\hline & $w$   & $\tau$   & $r$ &  $\theta$  & $p_r$  & $p_{\theta}$  \\
\hline
$H$ & /   & 1       & 11.0031 &  1.5867  & 7.6213e-3  & 1.9255 \\
$K$ & 1   & 1       & 11.0031 &  1.5867  & 7.6203e-3  & 1.9255  \\
\hline
$H$ & /   & 10      & 11.3074 & 1.7235   & 7.3287e-2  & 1.7922 \\
$K$ & 10  & 10.0002 & 11.3075 & 1.7235   & 7.3281e-2  & 1.7921  \\
\hline
$H$ & /   & $10^2$   & 26.2463 &  1.9136  & 0.2041  & -1.0078 \\
$K$ & $10^2$    & 100.0094   & 26.2473 & 1.9136   & 0.2041  & -1.0079  \\
\hline
$H$ & /   & $10^3$   & 126.4088 &  1.5798  & 6.3895e-2  & -1.8227 \\
$K$ & $10^3$    & 1000.0118   & 126.3912 & 1.5797   & 6.3868e-2  & -1.8226  \\
\hline
$H$ & /   & $10^4$   & 132.8861 &  1.5449  &  -5.1566e-2 & -3.0532 \\
$K$ & $10^4$    & 10000.1079   & 132.7308 & 1.5441   & -5.1730e-2  & -3.0572  \\
\hline
$H$ & /   & $10^5$   & 37.3004 &  1.1427  & -0.1763  & -1.4168 \\
$K$ & $10^5$    & 100001.0807   & 33.7413 & 1.1251   & -0.1827  & -1.2180  \\
\hline
$H$ & /   & $10^6$   & 150.9389 &  1.6606  &  1.2518e-2 & -2.0503 \\
$K$ & $10^6$    & 1000011.0250   & 151.8362 & 1.6616   & -2.7765e-3  & -2.0940  \\
\hline
$H$ & /   & $10^7$   & 142.0132 & 1.5745   & -3.5099e-2  & -3.0676 \\
$K$ & $10^7$    & 10000110.1522   & 110.0560 &  1.6668  & 8.0709e-2  & -3.0845  \\
\hline
\end{tabular}
\end{table*}

\begin{figure*}[ptb]
\center{
\includegraphics[scale=0.25]{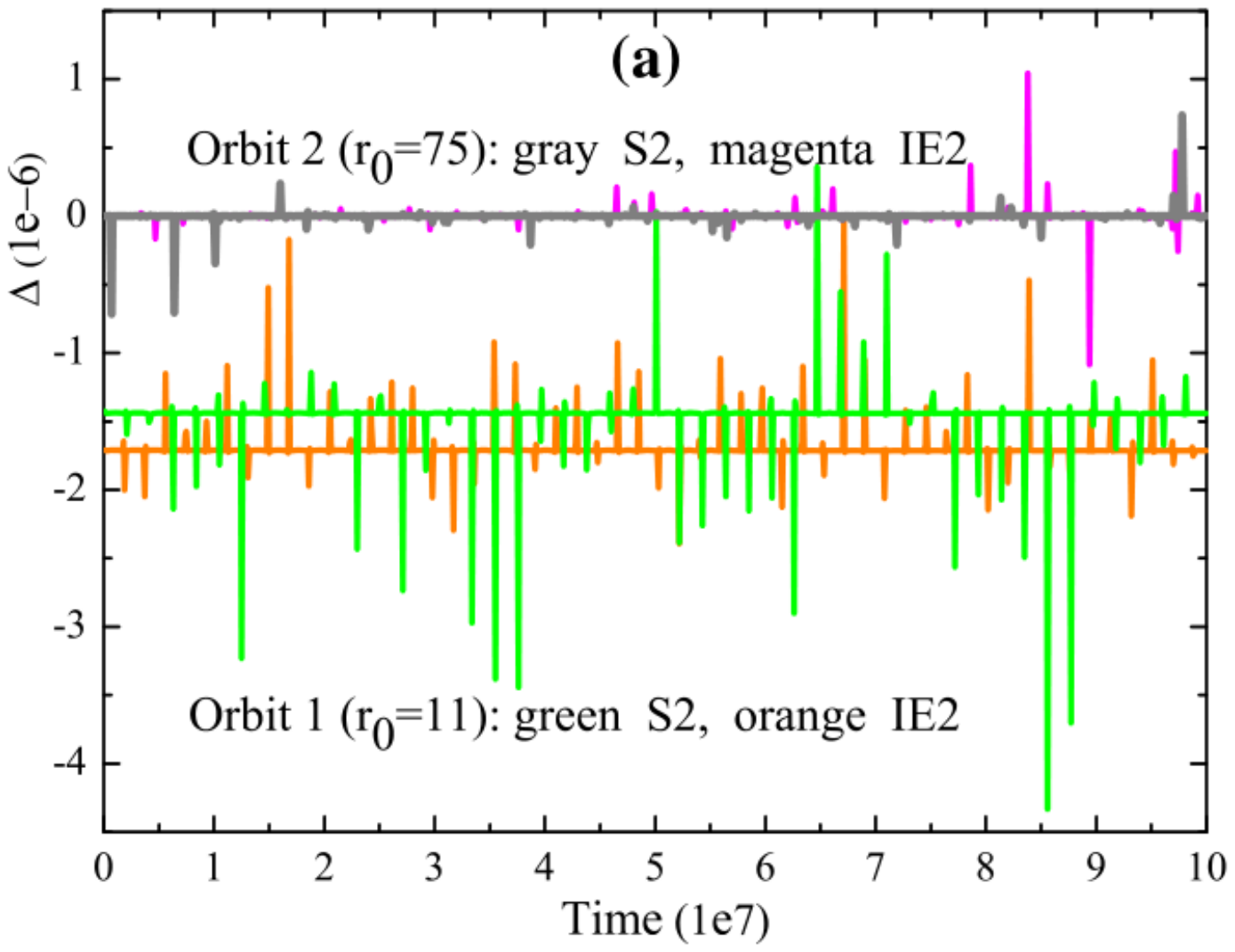}
\includegraphics[scale=0.25]{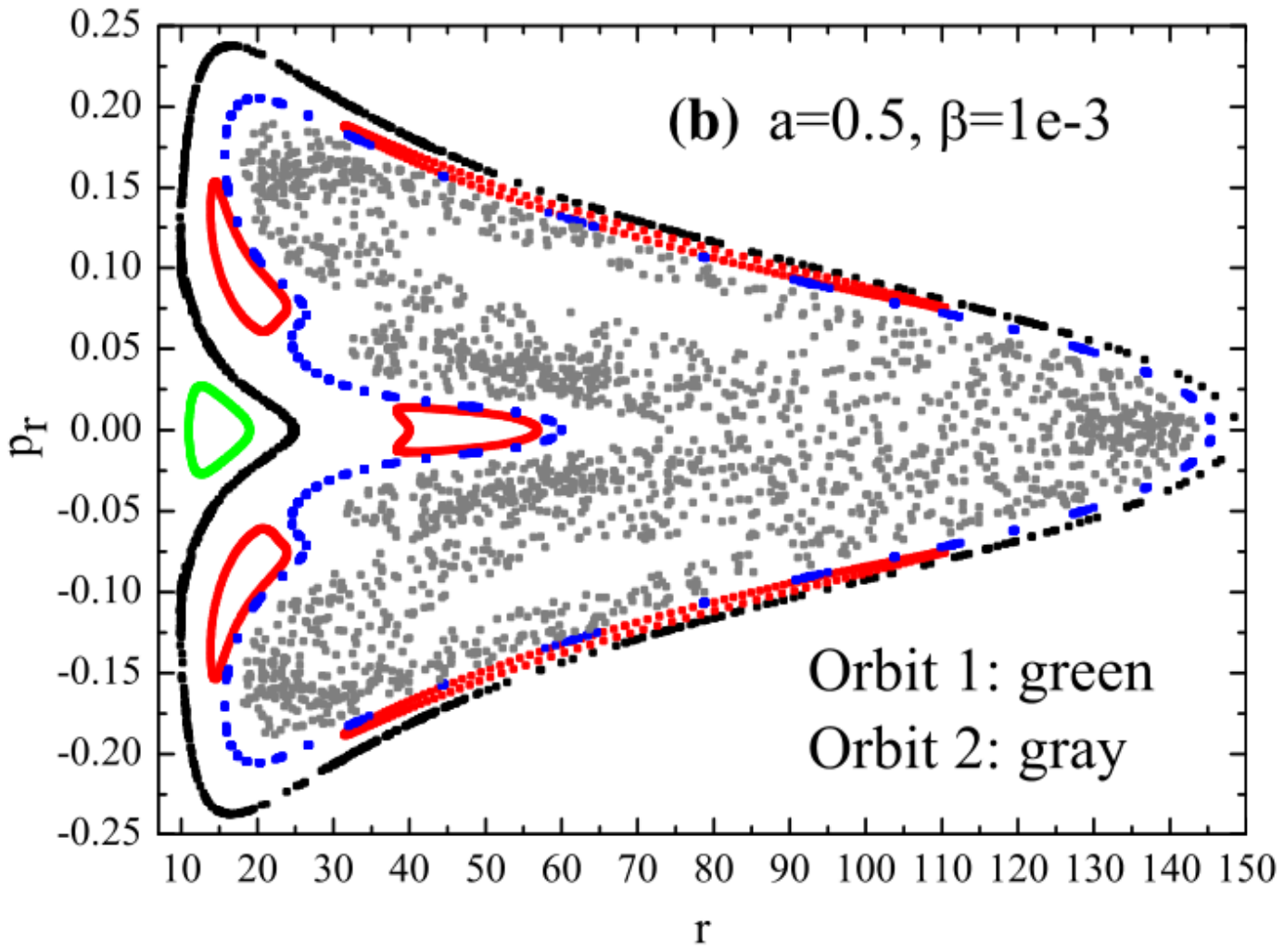}
\caption{(a) Hamiltonian errors $\Delta =K$ for the explicit
symplectic algorithm S2 and $\Delta =-H-1/2$ for the second-order
implicit and explicit  mixed symplectic method IE2. The parameters
are $E=0.995$, $L=4.6$, $a=0.5$ and $\beta=0.001$. Orbits 1 and 2
have their initial separations $r_0=11$ and 75, respectively. The
other initial conditions are $\theta=\pi/2$, $p_r=0$ and
$p_{\theta}>0$ given by Eqs. (14) and (16). $h=1$ is a proper time
step for IE2 and a new coordinate time step for S2. Both
algorithms make the errors have no secular drift and remain of the
same order for Orbits 1 and 2. (b) Poincar\'{e} sections on the
plane $\theta=\pi/2$ and $p_{\theta}>0$, given by S2. It is clear
that Orbit 1 is regular, whereas Orbit 2 is chaotic.}
 \label{Fig1}}
\end{figure*}

\begin{figure*}[ptb]
\center{
\includegraphics[scale=0.19]{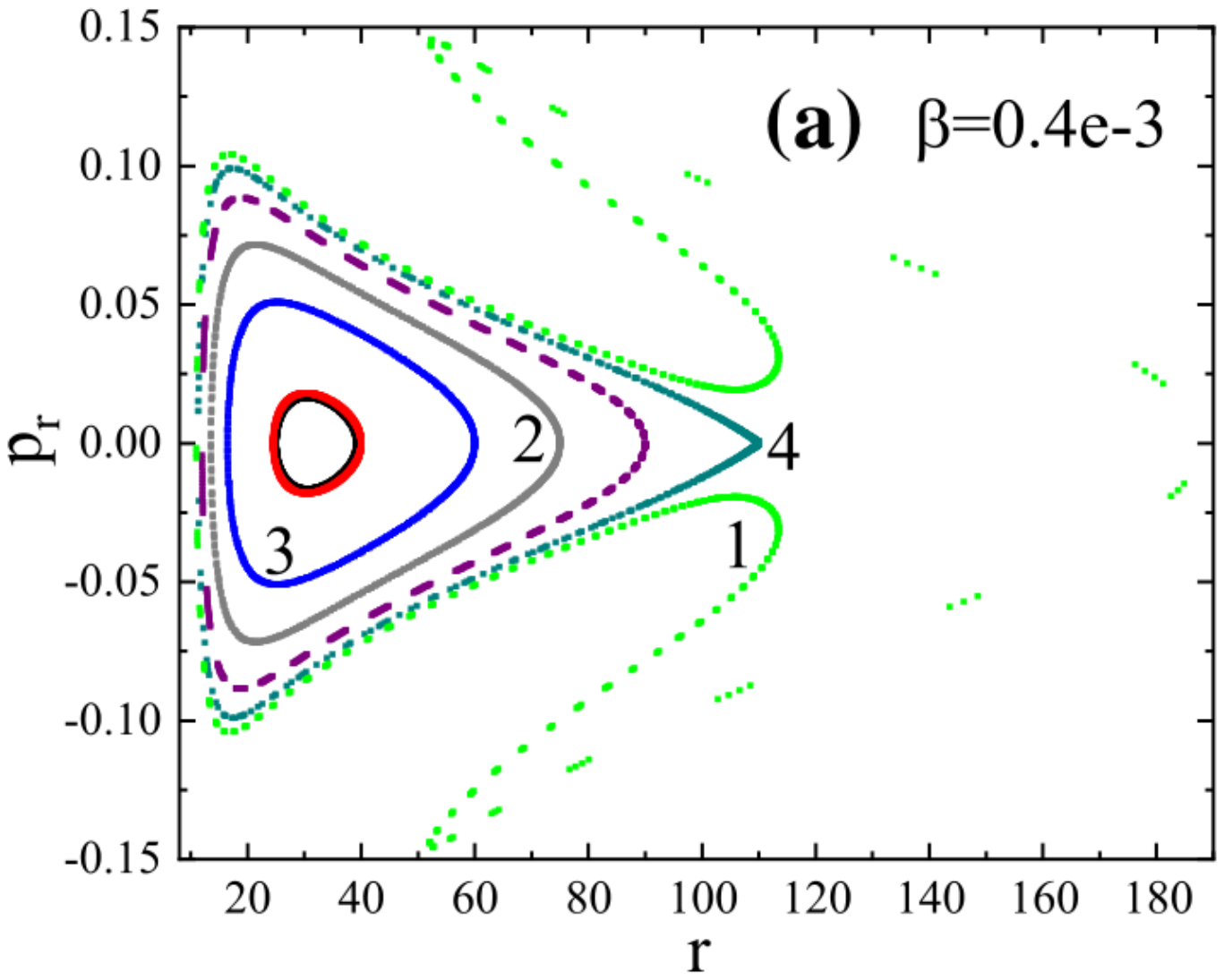}
\includegraphics[scale=0.19]{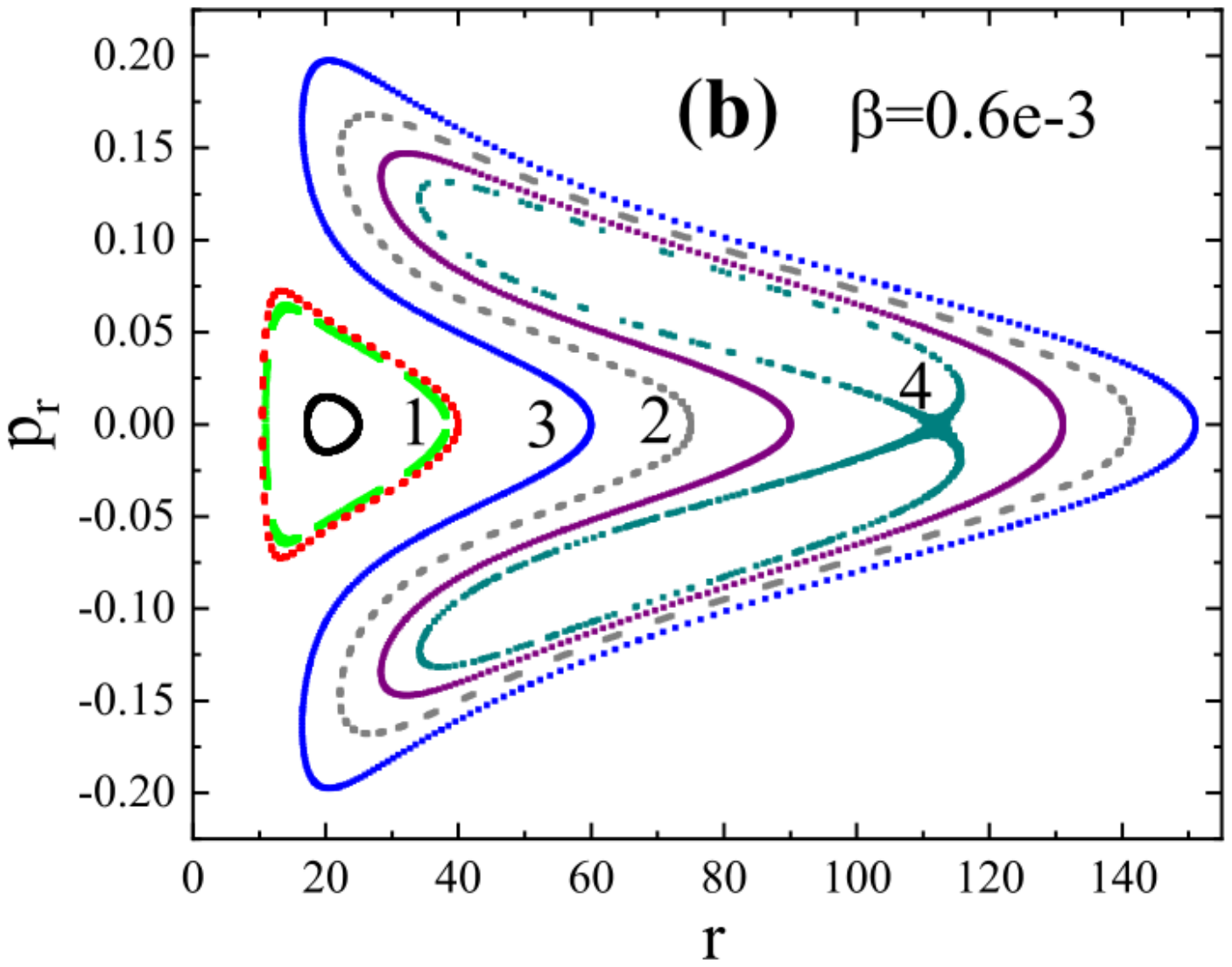}
\includegraphics[scale=0.19]{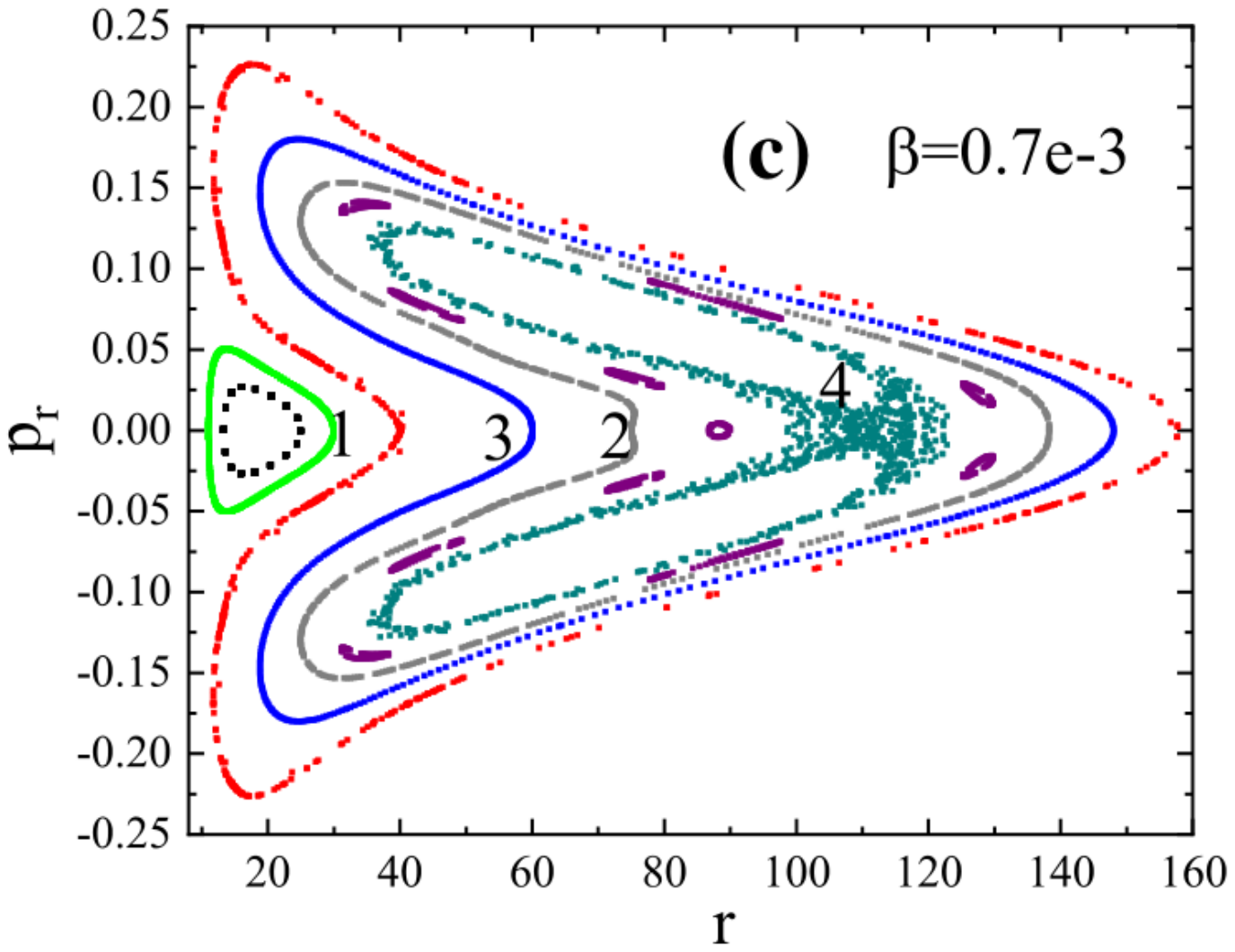}
\includegraphics[scale=0.19]{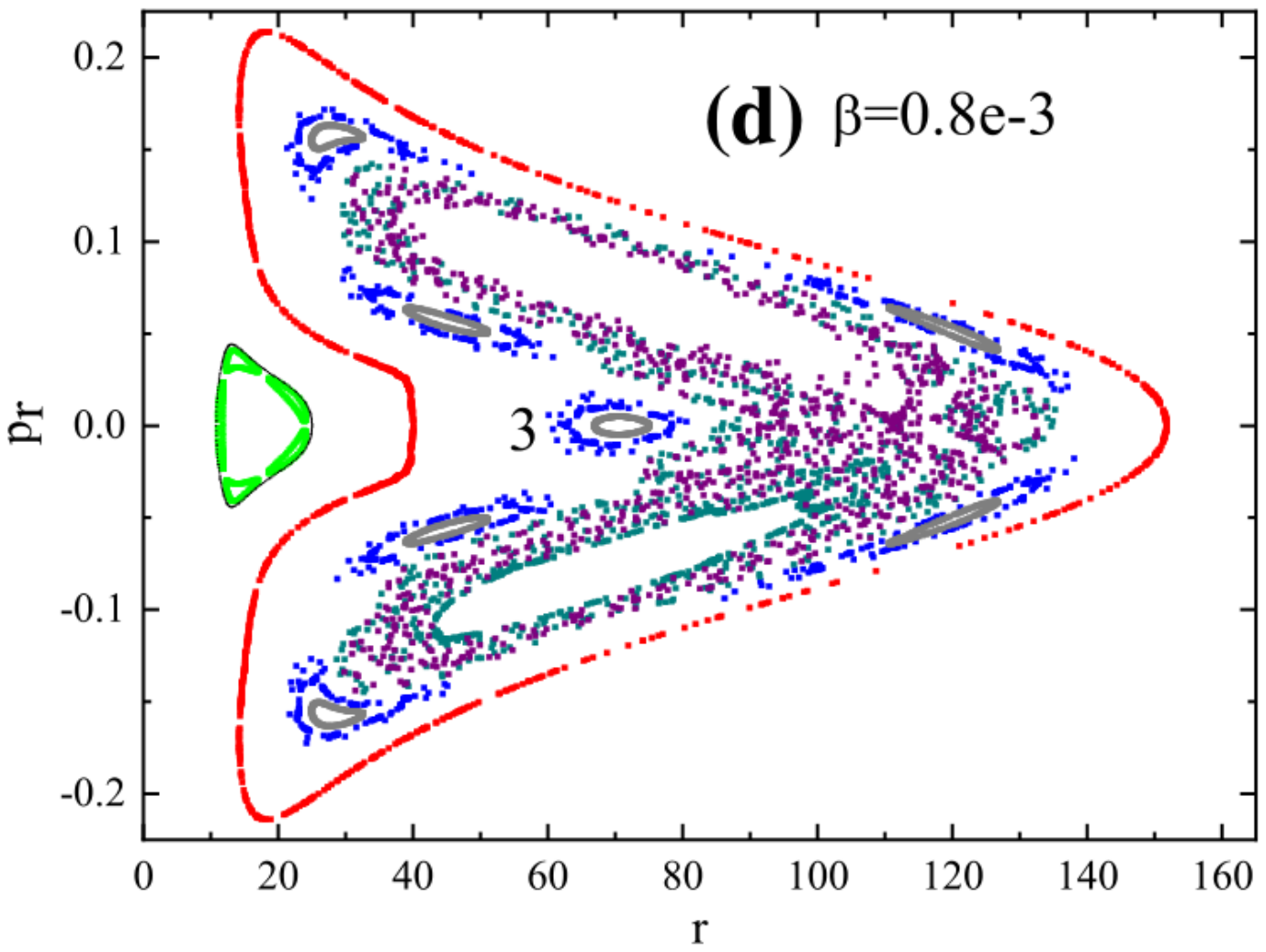}
\includegraphics[scale=0.19]{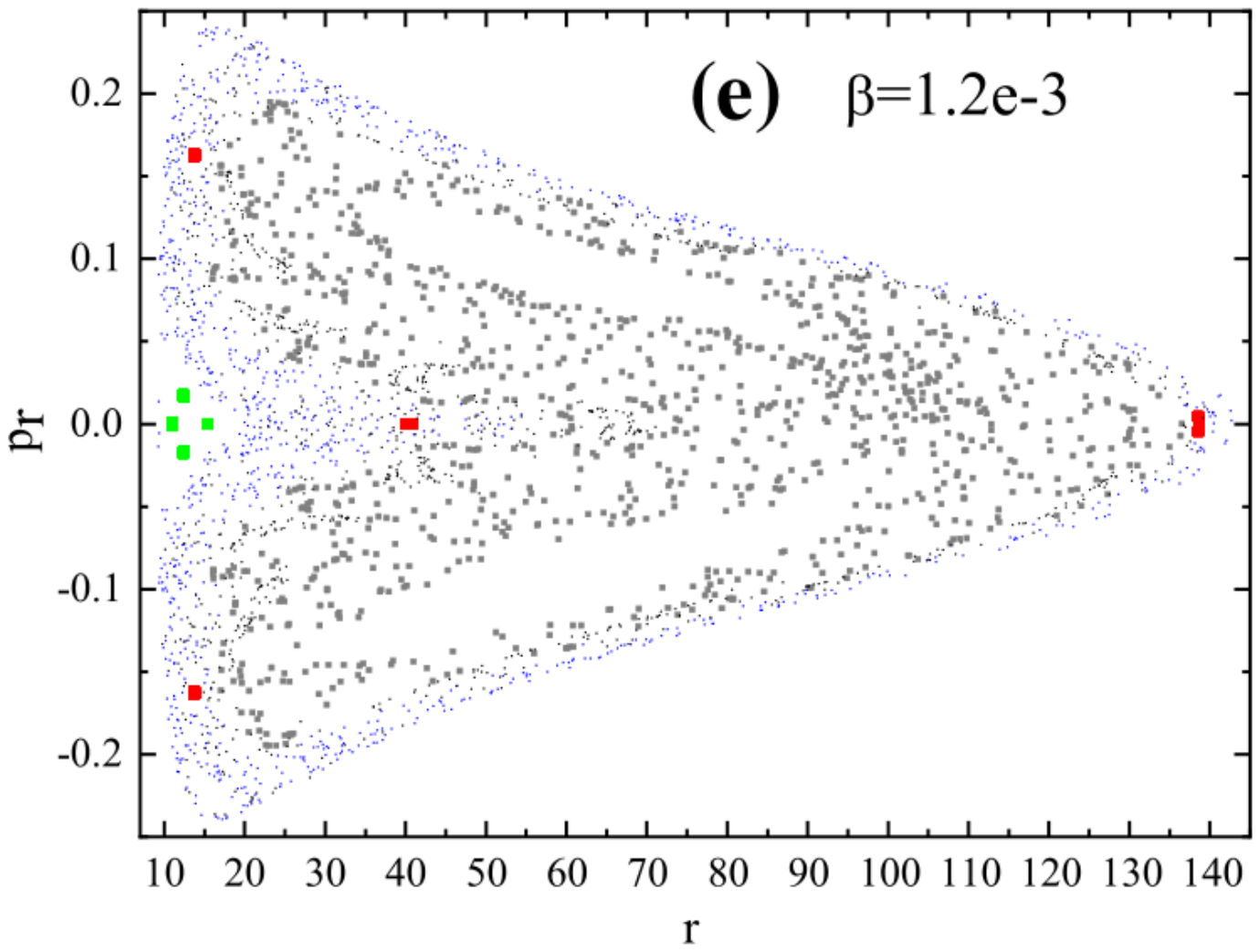}
\includegraphics[scale=0.19]{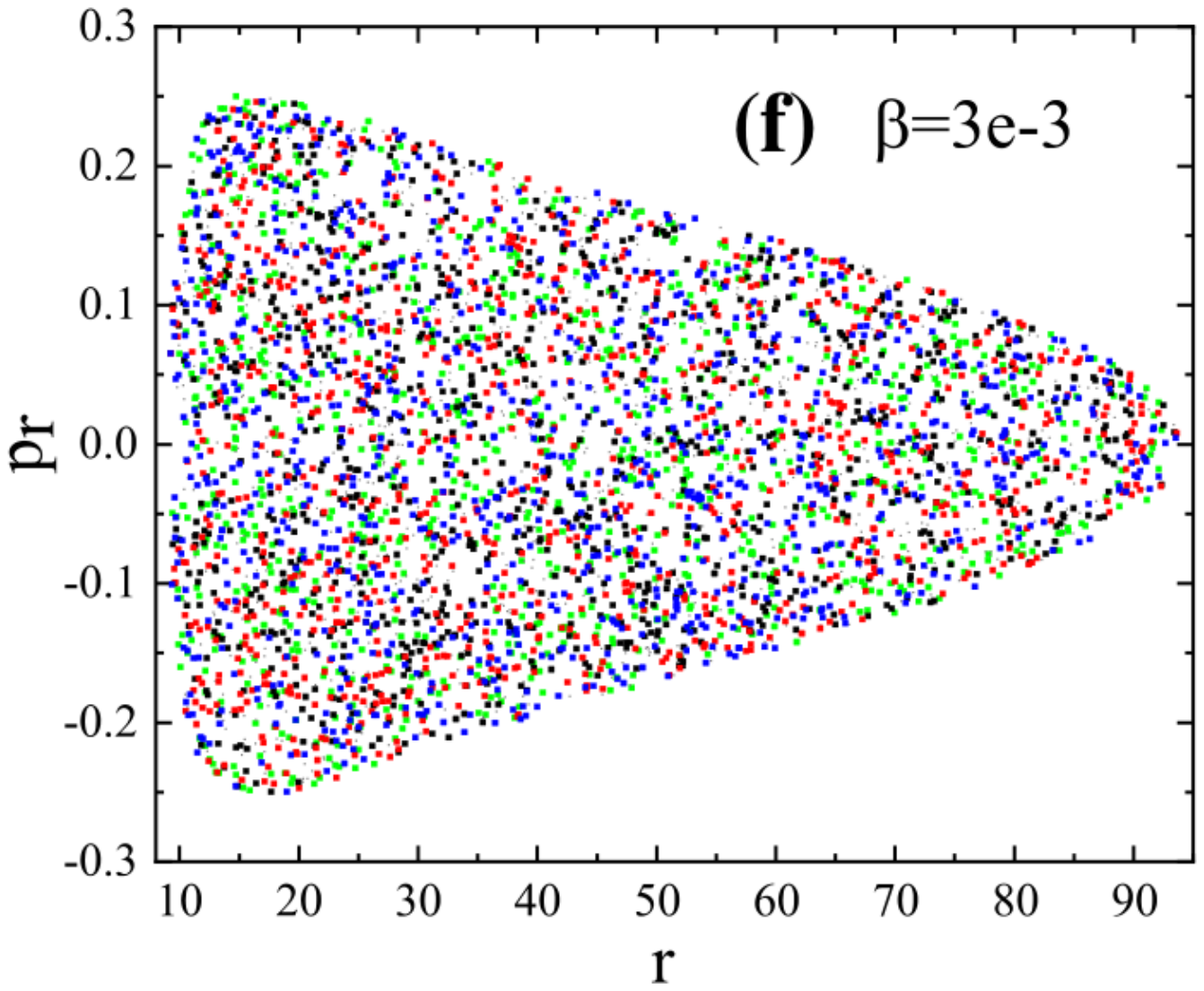}
\caption{Same as Fig. 1(b), but for different values of the
magnetic parameter $\beta$. An increase of $\beta$ results in
strengthening the extent of chaos. Orbit 3 with the initial
separation $r_0=60$ in panel (d) is weakly chaotic. }
 \label{Fig2}}
\end{figure*}

\begin{figure*}[ptb]
\center{
\includegraphics[scale=0.25]{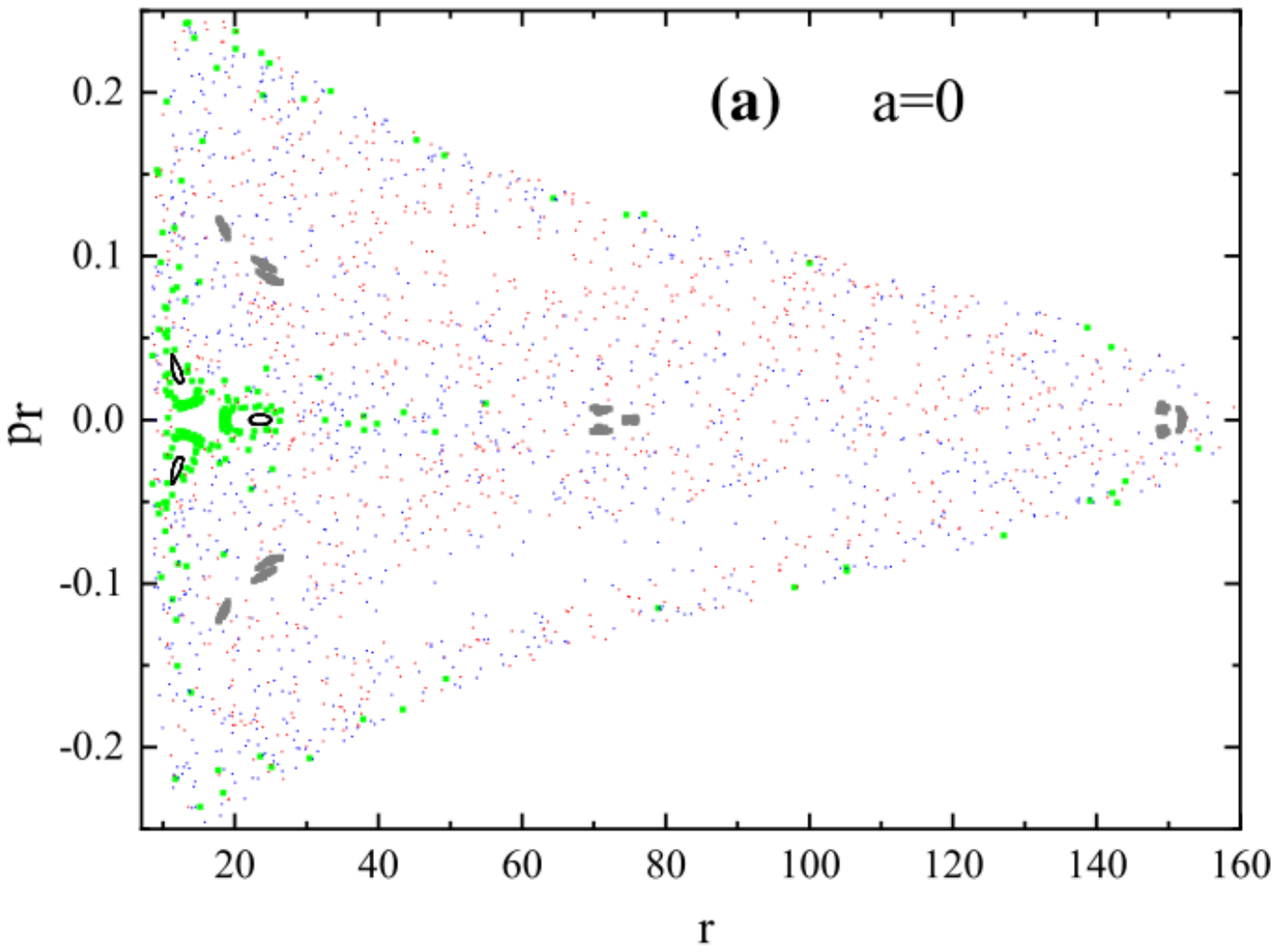}
\includegraphics[scale=0.25]{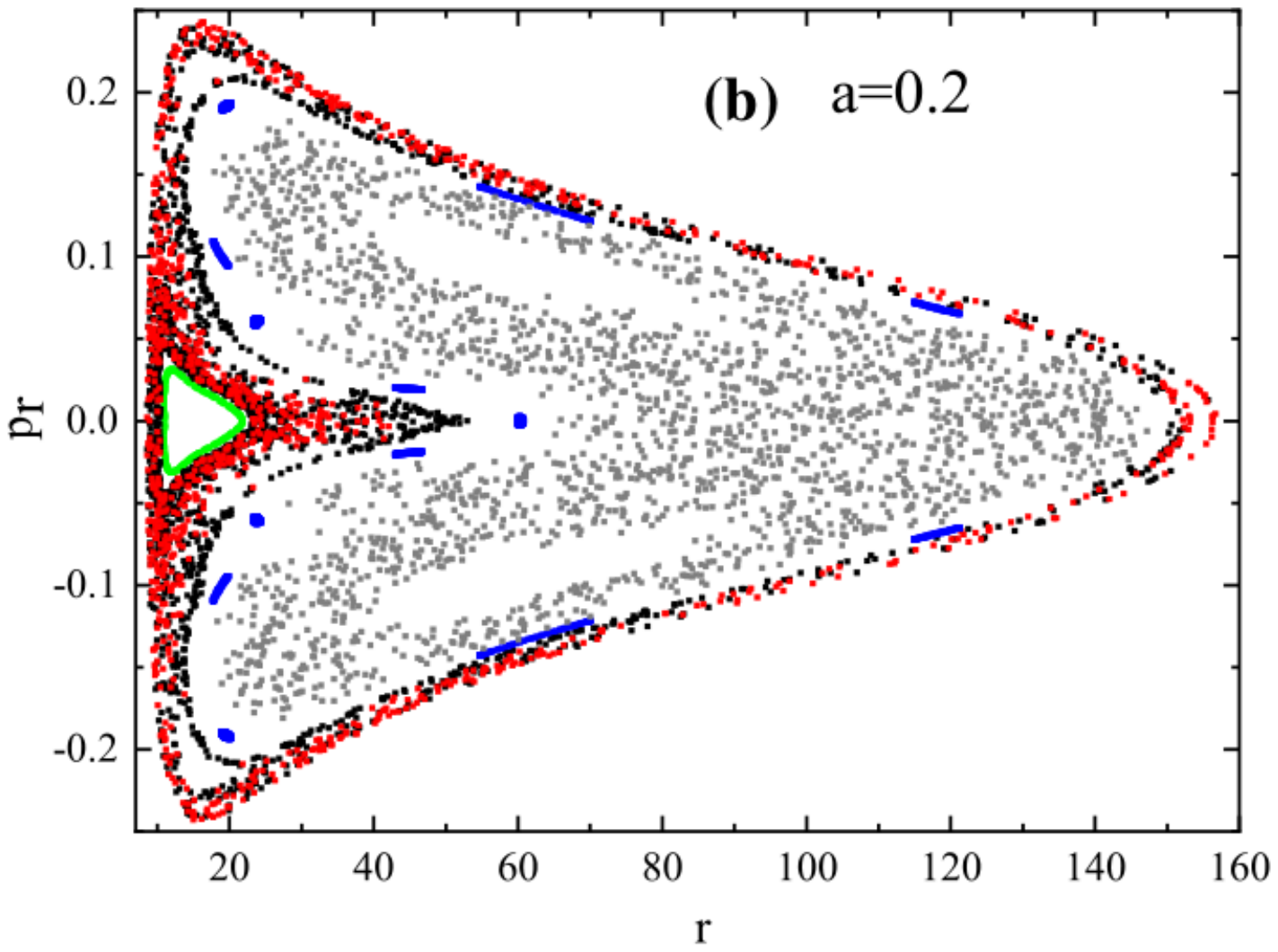}
\includegraphics[scale=0.25]{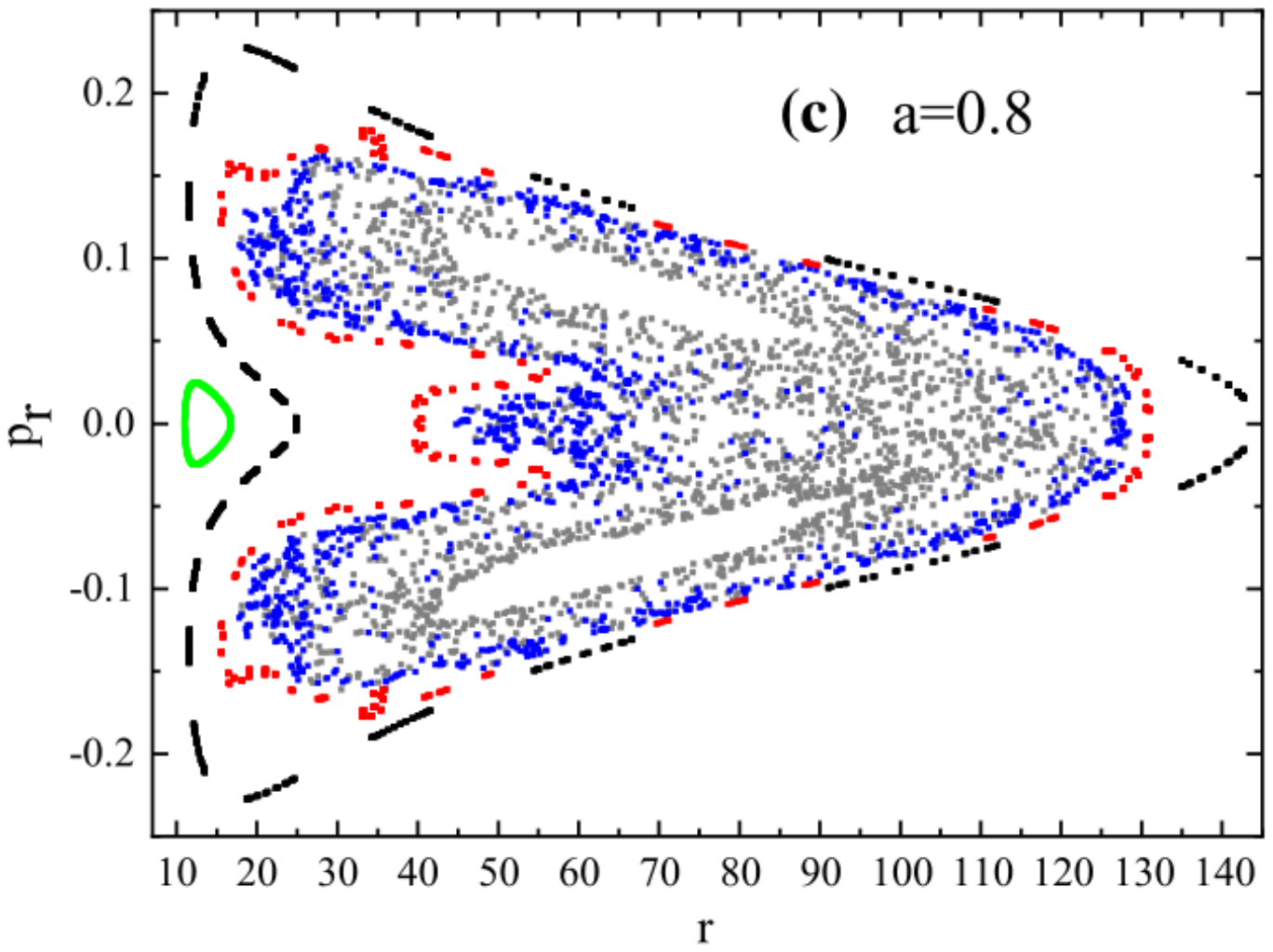}
\includegraphics[scale=0.25]{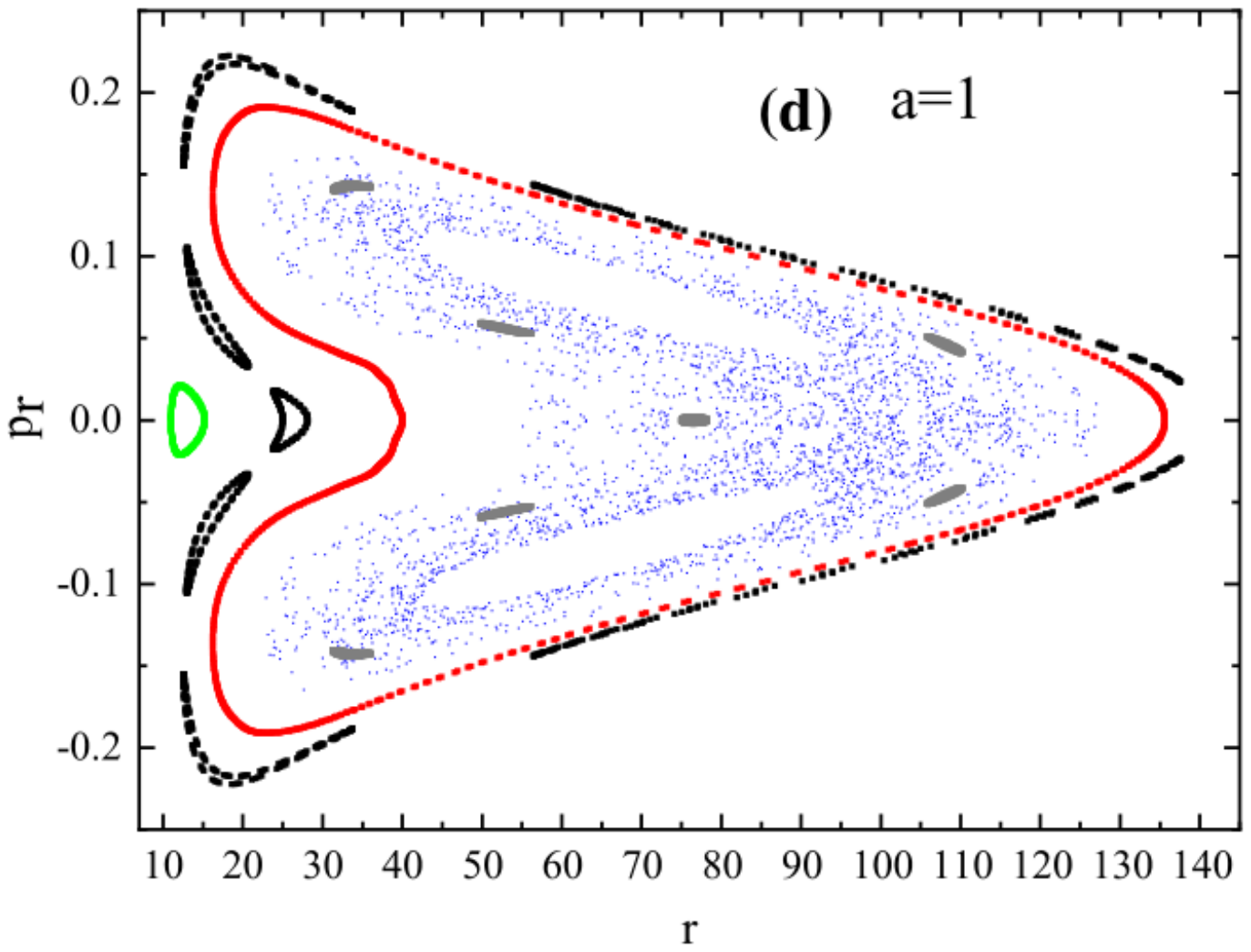}
\caption{Same as Fig. 1(b), but for different values of the black
hole's angular momentum $a$. Here, the magnetic parameter
$\beta=1\times 10^{-3}$. An increase of $a$ seems to weaken the
extent of chaos from the global phase space structure.}
 \label{Fig3}}
\end{figure*}

\begin{figure*}[ptb]
\center{
\includegraphics[scale=0.25]{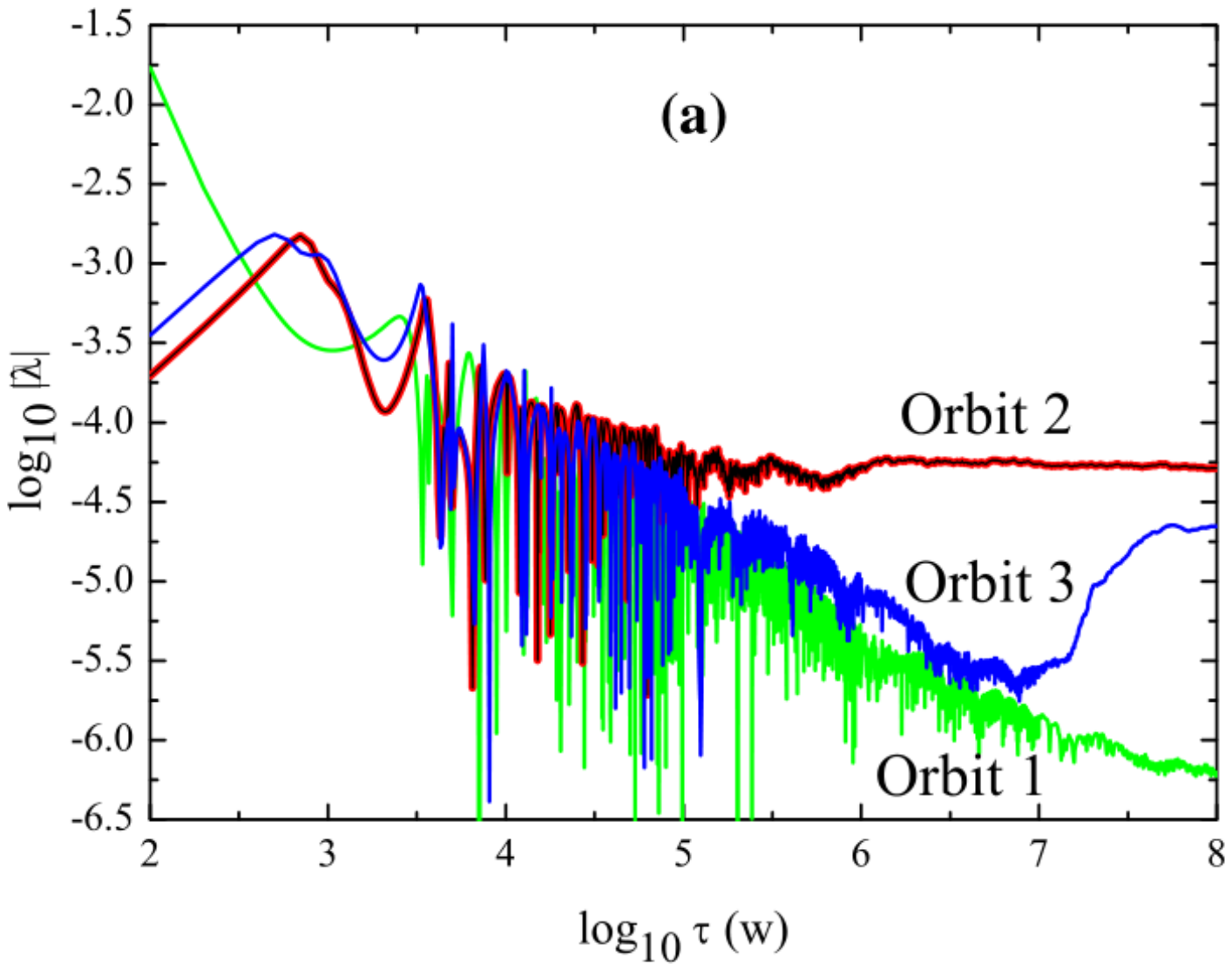}
\includegraphics[scale=0.25]{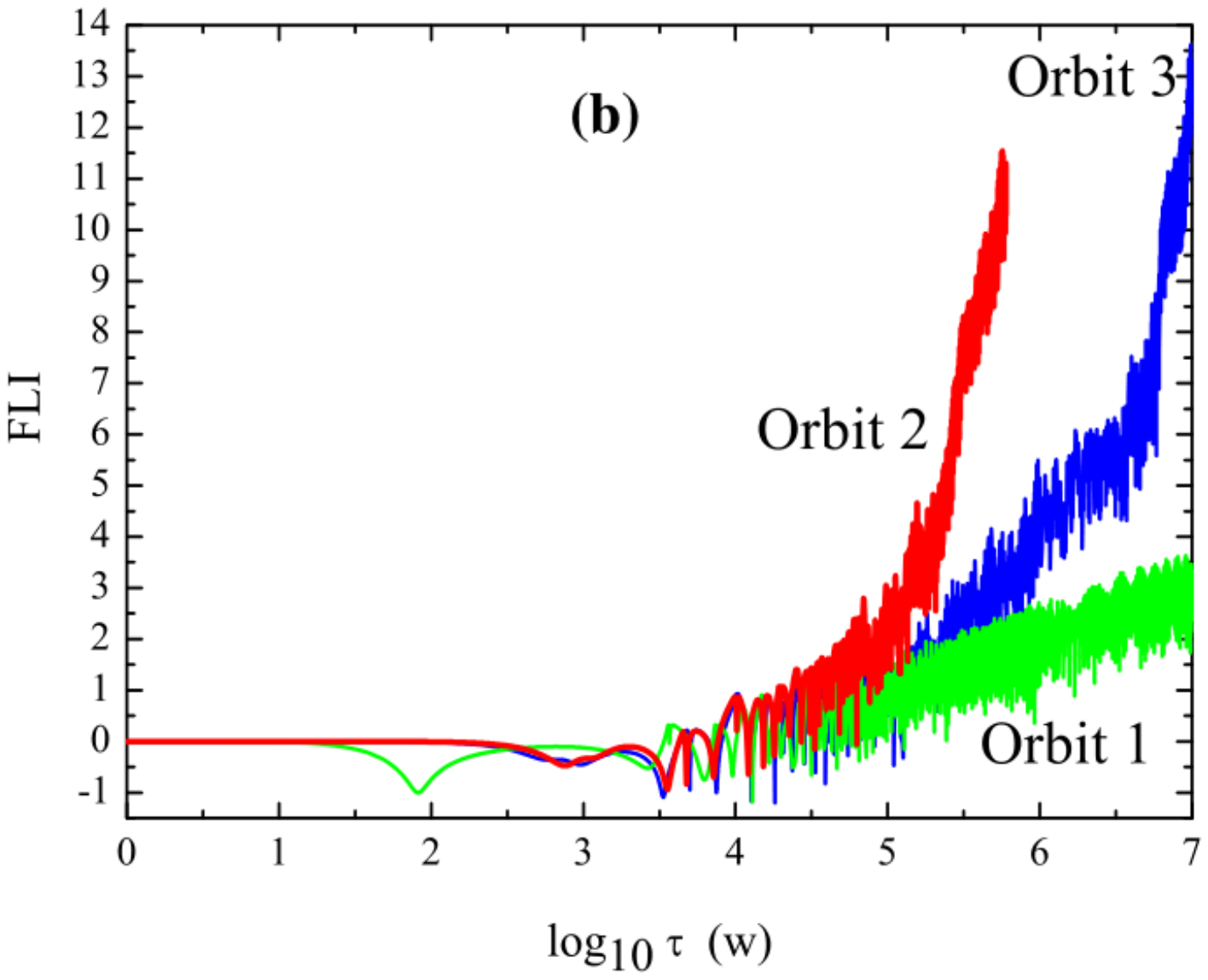}
\caption{(a) Lyapunov exponents of three orbits including Orbits 1
and 2 in Fig. 1 and Orbit 3 in Fig. 2(d). The Lyapunov exponent
(colored red) of Orbit 2 calculated in the new coordinate time $w$
coincides with that (colored black) of Orbit 2 computed in the
proper time $\tau$. Orbit 1 is regular, whereas Orbits 2 and 3
show the onset of chaos. (b) Fast Lyapunov indicators (FLIs) of
the three orbits. The FLI of Orbit 2 reaches 12 when the
integration time $w=6\times10^5$, but is larger than 200 for the
integration time $w=10^7$. Thus, this orbit is strongly chaotic.
Weak chaos of Orbit 3 can be distinguished from regular Orbit 1 in
terms of the growth of FLIs of the two orbits during the
integration time $w=10^7$.}
 \label{Fig4}}
\end{figure*}

\begin{figure*}[ptb]
\center{
\includegraphics[scale=0.25]{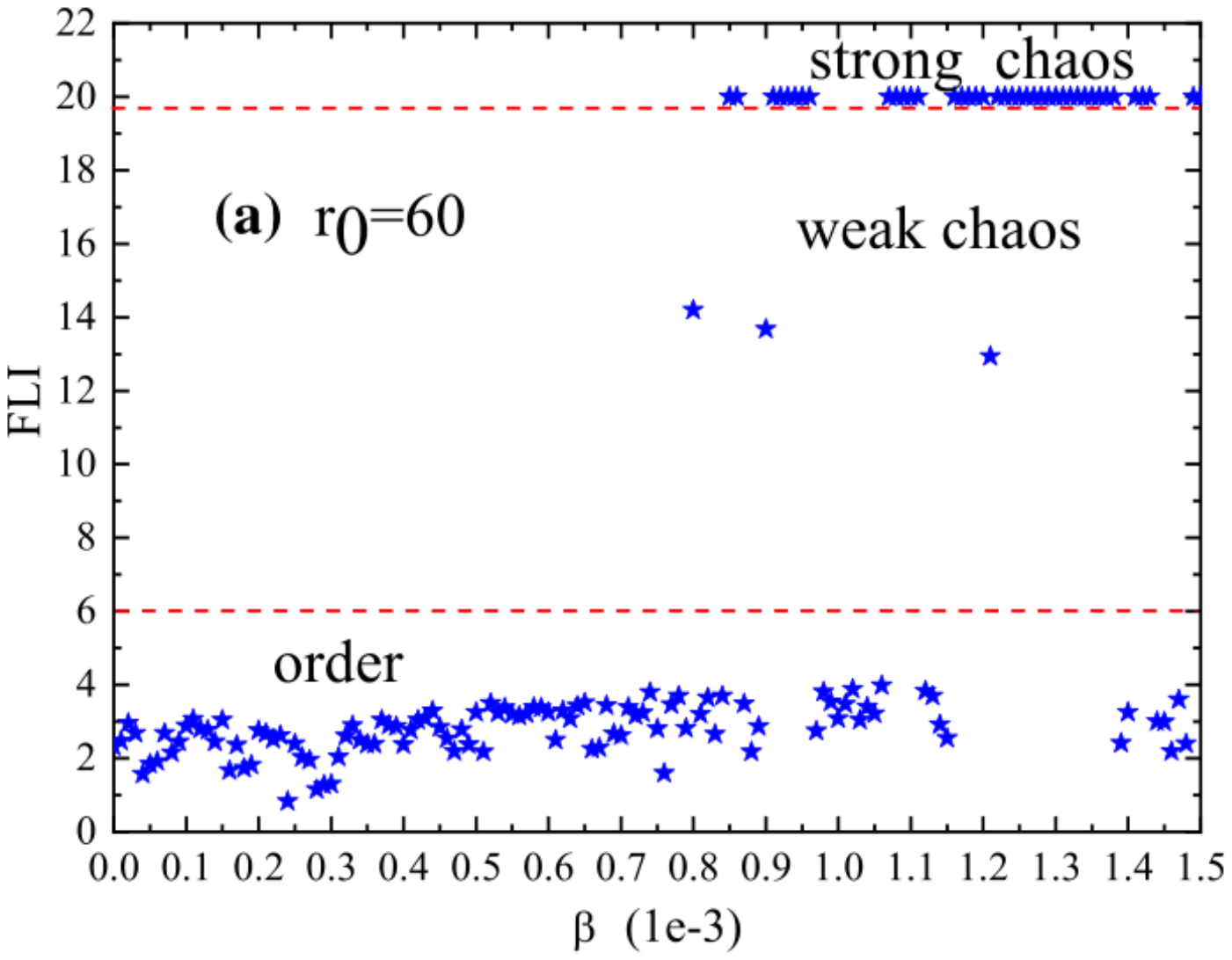}
\includegraphics[scale=0.25]{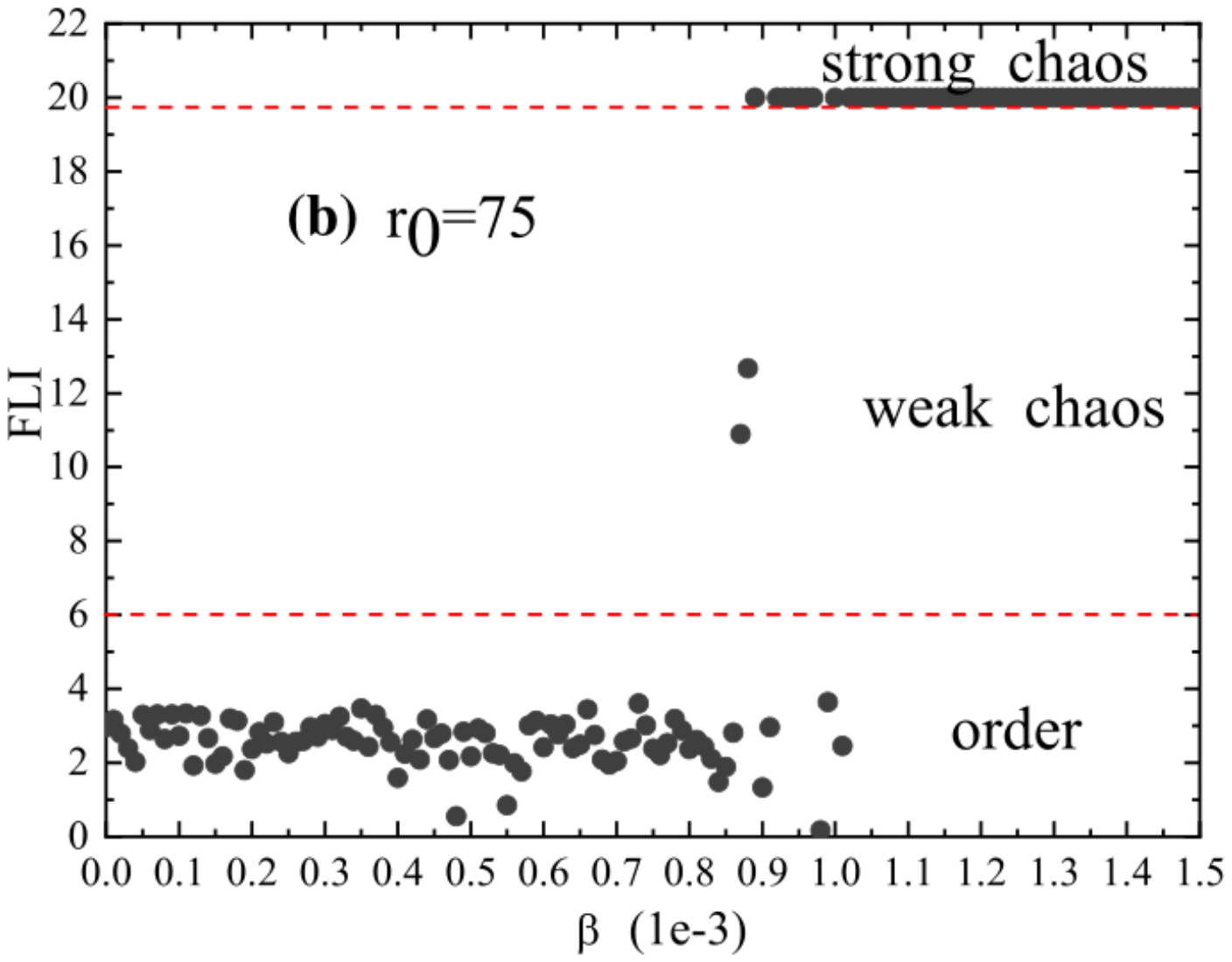}
\caption{Dependence of the values of FLIs on the magnetic
parameter $\beta$. The other parameters are $E=0.995$, $L=4.6$ and
$a=0.5$. The initial conditions $\theta=\pi/2$ and $p_r=0$ are the
same in the two panels, but the initial values $r_0$ and
$p_{\theta}>0$ are different. For the strong chaotic case, the
integration ends when the FLI reaches 20, whereas the integration
time $w=10^{7}$ for the two other cases.}
 \label{Fig5}}
\end{figure*}

\begin{figure*}[ptb]
\center{
\includegraphics[scale=0.25]{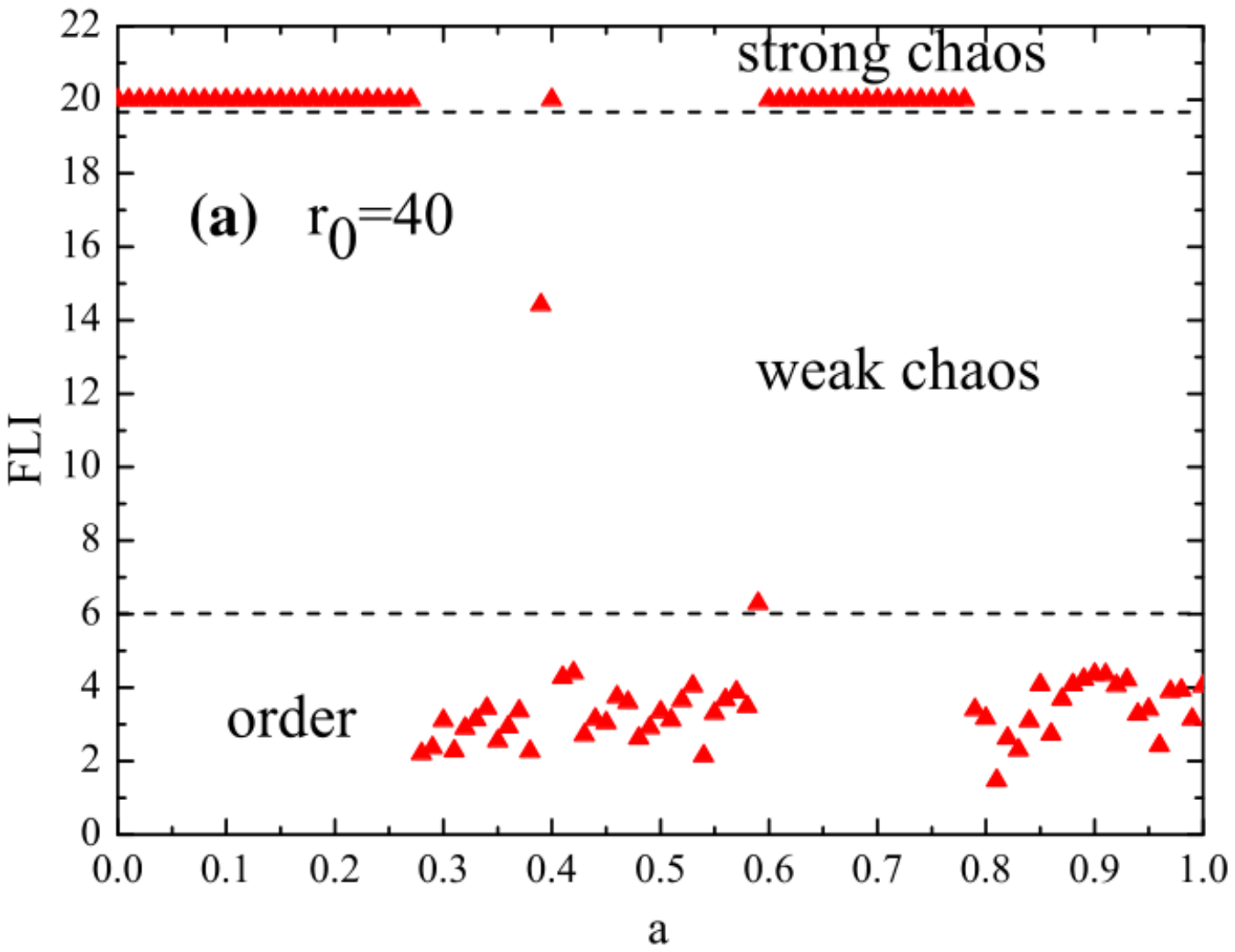}
\includegraphics[scale=0.25]{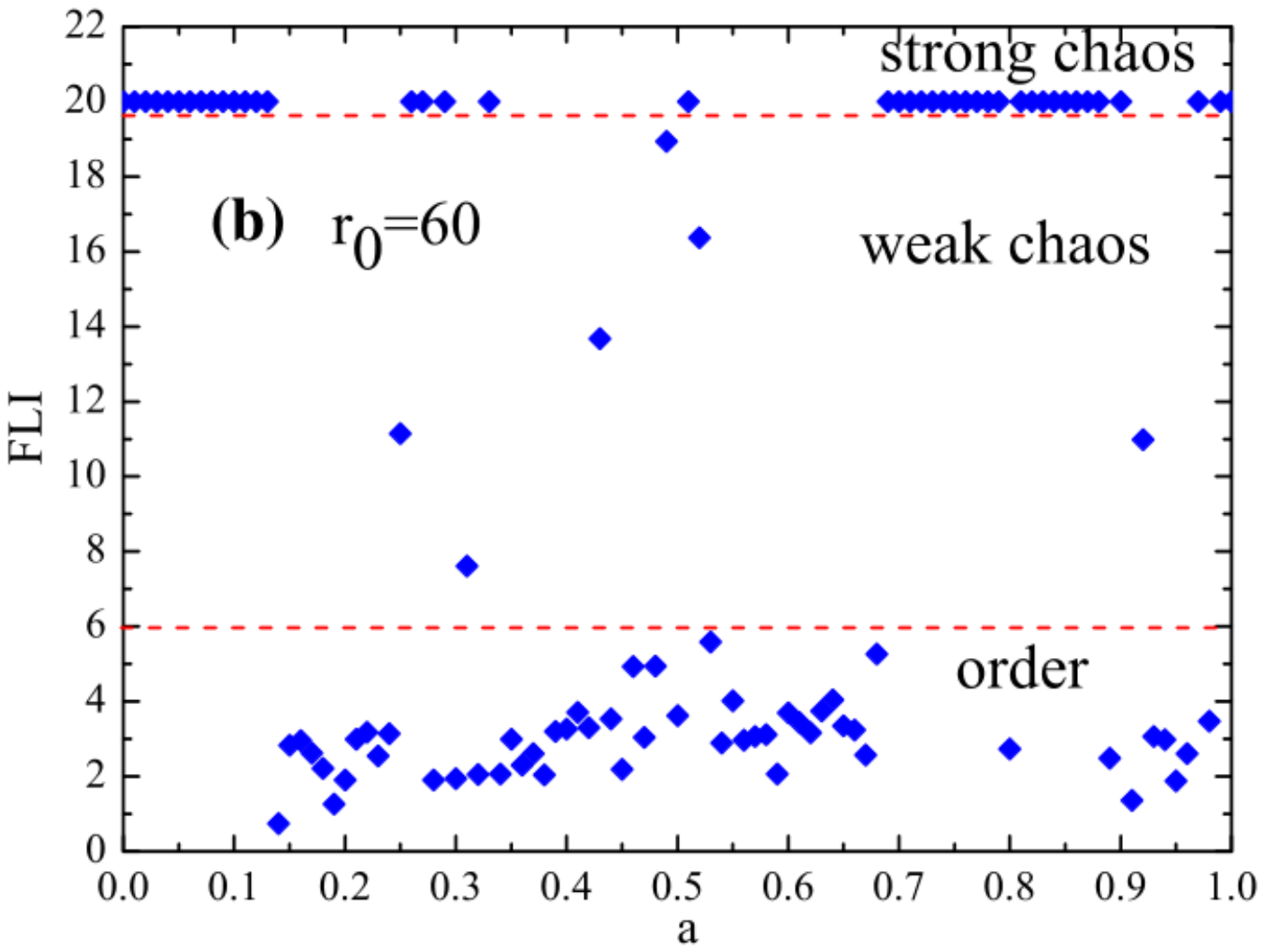}
\includegraphics[scale=0.25]{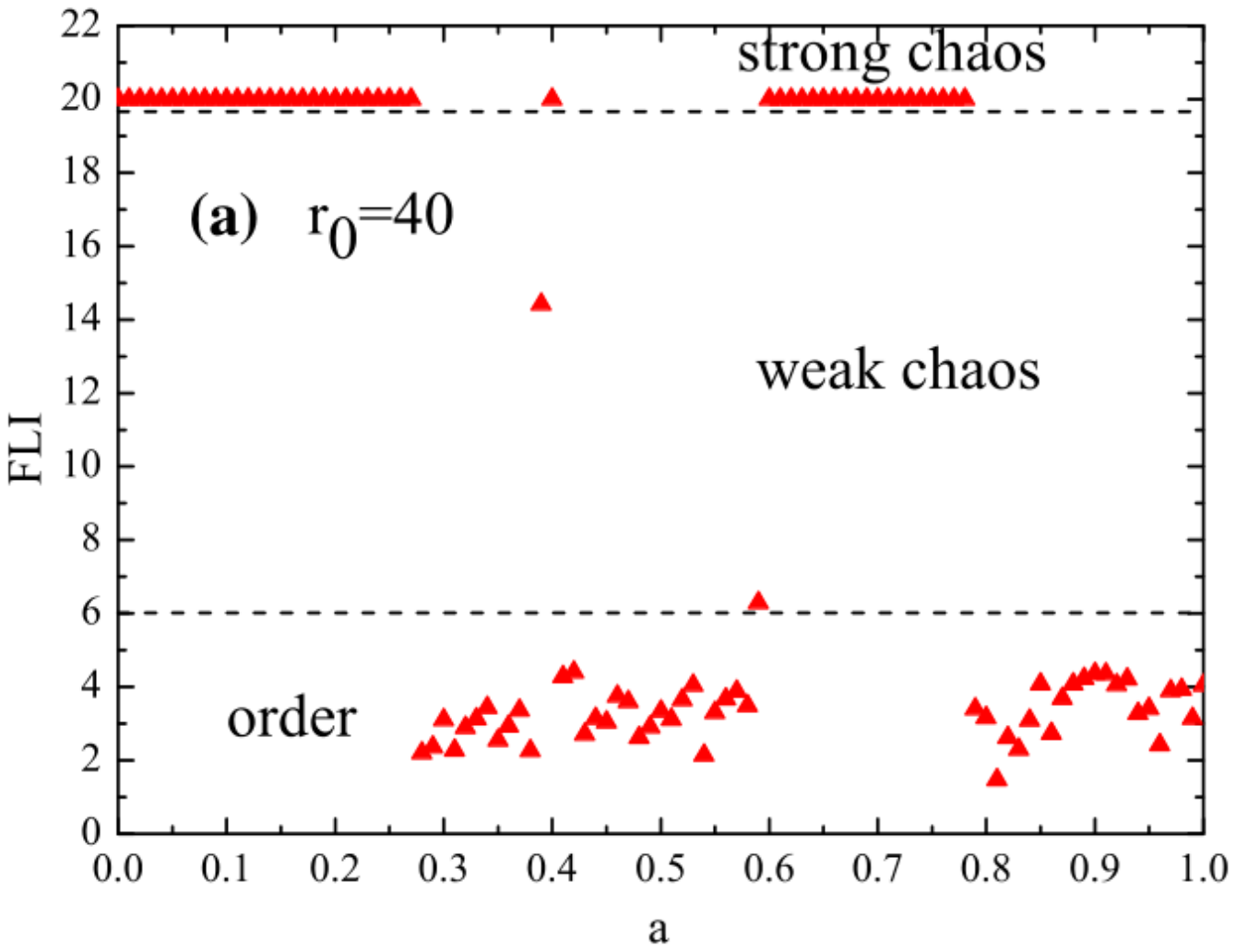}
\includegraphics[scale=0.25]{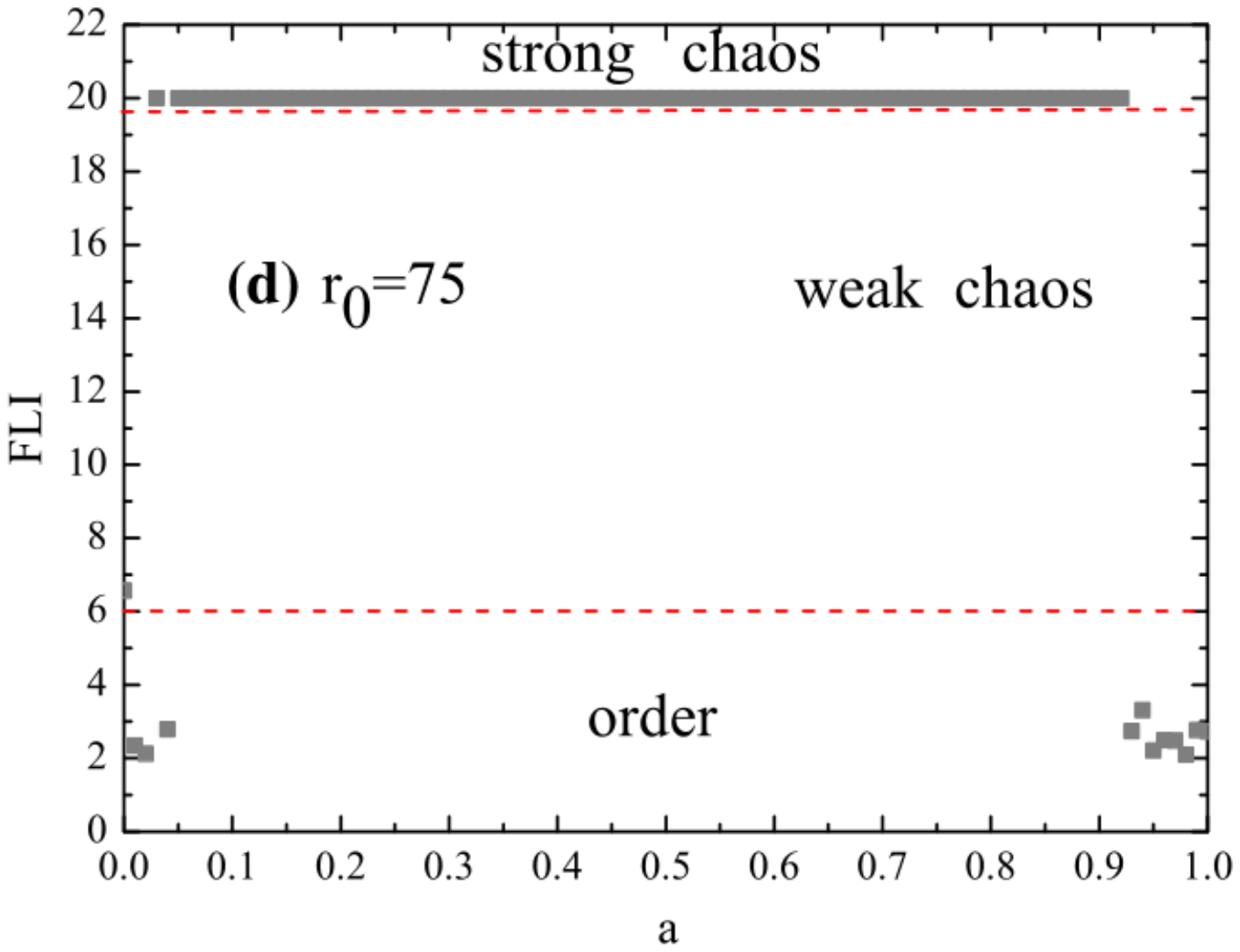}
\caption{Same as Fig. 5, but the dependence of the values of FLIs
on the black hole spin $a$. Here, $\beta=0.001$. }
 \label{Fig6}}
\end{figure*}


\begin{thebibliography}{99}

\bibitem[Kerr(1963)]{1} R. P. Kerr, Phy. Rev. Lett. \textbf{11}, 237 (1963)

\bibitem[Carter(1968)]{2} B. Carter, Phy. Rev. \textbf{174}, 1559 (1968)

\bibitem[Nakamura \& Ishizuka(1968)]{3} Y. Nakamura, T. Ishizuka, Astrophys. Space Science \textbf{210}, 105
(1993)

\bibitem[Kop\'{a}\v{c}ek et al.(2010)]{4} O. Kop\'{a}\v{c}ek, V. Karas, J. Kov\'{a}\v{r}, Z.
Stuchl\'{i}k, Astrophys. J. \textbf{722}, 1240 (2010)

\bibitem[Kop\'{a}\v{c}ek \& Karas (2014)]{5} O. Kop\'{a}\v{c}ek, V. Karas, Astrophys. J. \textbf{787},
117 (2014)

\bibitem[Kop\'{a}\v{c}ek \& Karas (2018)]{6} O. Kop\'{a}\v{c}ek, V. Karas, Astrophys. J. \textbf{853}, 53
(2018)

\bibitem[Frolov \& Shoom(2010)]{7} V. P. Frolov, A. A. Shoom, Phys. Rev.
D \textbf{82}, 084034 (2010)

\bibitem[Kop\'{a}\v{c}ek et al.(2017)]{8} M. Kolo\v{s}, A. Tursunov, Z.
Stuchl\'{i}k, Eur. Phys. J. C \textbf{77}, 860 (2017)

\bibitem[P\'{a}nis et al.(2019)]{9} R. P\'{a}nis, M. Kolo\v{s}, Z.
Stuchl\'{i}k, Eur. Phys. J. C \textbf{79}, 479 (2019)

\bibitem[Stuchl\'{i}k et al.(2020)]{10} Z. Stuchl\'{i}k, M. Kolo\v{s}, J.
Kov\'{a}\v{r}, P. Slan\'{y}, A. Tursunov, Universe \textbf{6}, 26
(2020)

\bibitem[Takahashi \& Koyama(2009)]{11} M. Takahashi, H. Koyama, Astrophys. J. \textbf{693}, 472
(2009)

\bibitem[Wisdom(1982)]{12} J. Wisdom, Astron. J. \textbf{87}, 577 (1982)

\bibitem[Ruth(1983)]{13} R. D. Ruth, IEEE Trans. Nucl. Sci. NS \textbf{30}, 2669 (1983)

\bibitem[Wisdom \& Holman(1991)]{14} J. Wisdom J., M. Holman, Astron. J. 102, 1528 (1991)

\bibitem[Feng(1986)]{15} K. Feng, Journal of Computational Mathematics \textbf{44},
279 (1986)

\bibitem[Brown(2006)]{16} J. D. Brown, Phys. Rev. D \textbf{73}, 024001 (2006)

\bibitem[Liao(2006)]{17} X. H. Liao, Celest. Mech. Dyn. Astron. \textbf{66},
243 (1997)

\bibitem[Preto(2009)]{18} M. Preto, P. Saha, Astrophys. J. \textbf{703}, 1743
(2009)

\bibitem[Lubich \& Walther(2010)]{19} C. Lubich, B. Walther, B. Br\"{u}gmann, Phys. Rev. D
\textbf{81}, 104025 (2010)

\bibitem[Zhong et al.(2010)]{20} S. Y. Zhong, X. Wu, S. Q. Liu, X. F. Deng, Phys. Rev. D \textbf{82},
124040 (2010)

\bibitem[Mei et al.(2013a)]{21} L. Mei, X. Wu, F. Liu, Eur. Phys. J. C \textbf{73},
2413 (2013)

\bibitem[Mei et al.(2013b)]{22} L. Mei, M. Ju, X. Wu, S. Liu, Mon. Not. R. Astron. Soc. \textbf{435},
2246 (2013)

\bibitem[Wang et al.(2021a)]{23} Y. Wang, W. Sun, F. Liu, X. Wu, Astrophys. J. \textbf{907}, 66 (2021) (Paper I)

\bibitem[Wang et al.(2021b)]{24} Y. Wang, W. Sun, F. Liu, X. Wu,  Astrophys. J. \textbf{909}, 22 (2021) (Paper II)

\bibitem[Wang et al.(2021c)]{25} Y. Wang, W. Sun, F. Liu, X. Wu, Astrophys. J. Supp. Series  \textbf{254}, 8 (2021) (Paper III)

\bibitem[Mikkola (1997)]{26} S. Mikkola, Celest. Mech. Dyn. Ast. \textbf{67}, 145
(1997)

\bibitem[Wu et al.(2021)]{27} X. Wu, Y. Wang, W. Sun, F. Liu, Astrophys. J.  \textbf{914},
63 (2021) (Paper IV)

\bibitem[Tancredi et al.(2001)] {28} G. Tancredi, A. S\'{a}nchez, F. Roig, Astron. J. \textbf{121},
1171 (2001)

\bibitem[Wu \& Huang(2003)]{29} X. Wu, T.-Y. Huang, Phys. Lett. A \textbf{313}, 77
(2003)

\bibitem[Froeschl\'{e} et al. (1997)]{30} C. Froeschl\'{e}, E. Lega, R. Gonczi, Celest. Mech. Dyn. Astron.
\textbf{67}, 41 (1997)

\bibitem[Froeschl\'{e} \& Lega (1997)]{31} C. Froeschl\'{e}, E. Lega, Celest. Mech. Dyn. Astron. \textbf{78},
167 (2000)

\bibitem[Wu et al.(2006)]{32} X. Wu, T.-Y. Huang, H. Zhang, Phys. Rev. D \textbf{74},
083001 (2006)

\bibitem[Stuchl\'{i}k \& Kolo\v{s} (2016)]{33} Z. Stuchl\'{i}k, M. Kolo\v{s}, Eur. Phys. J. C \textbf{76},
32 (2016)

\bibitem[Tursunov et al.(2016)]{34} A. Tursunov, Z. Stuchl\'{i}k, M. Kolo\v{s},
Phys. Rev D \textbf{93}, 084012 (2016)

\bibitem[Mikkola \& Tanikawa(1999)]{35} S. Mikkola, K. Tanikawa, Celest. Mech. Dyn. Ast. \textbf{74},
287 (1999)

\bibitem[Preto \& Tremaine(1999)]{36} M. Preto, S. Tremaine, Astron. J. \textbf{118}, 2532
(1999)

\end{thebibliography}
\end{document}